\documentclass[preprint,5p,twocolumn]{elsarticle}
\pdfoutput=1
\usepackage{hyperref,amsmath,amssymb,amsfonts,gensymb,comment,stackengine,graphicx,xcolor,lmodern}
\usepackage[T1]{fontenc}

\journal{Astroparticle Physics}
\newcommand{\dedx}{\ensuremath{\textrm{d}E/\textrm{d}x} }
\newcommand{\range}{\ensuremath{0.07-0.21}}
\newcommand{\totalTOI}{\textasciitilde600}
\newcommand{\totalcosmic}{\textasciitilde500}
\newcommand{\totalTOIProgram}{\textasciitilde1700}
\newcommand{\totalcosmicProgram}{\textasciitilde1500}

\begin{document}
\begin{frontmatter}

\title{Sensitivity of the GAPS Experiment to Low-energy Cosmic-ray Antiprotons}

\author[MIT,SSL]{F.~Rogers}
\corref{mycorrespondingauthor}
\cortext[mycorrespondingauthor]{Corresponding author}
\ead{frrogers@mit.edu}

\author[Northeastern]{T.~Aramaki}%
\author[INFN-trieste,IFPU]{M.~Boezio}%
\author[UCSD]{S.~E.~Boggs}%
\author[INFN-trieste]{V.~Bonvicini}%
\author[CU]{G.~Bridges}%
\author[INFN-napoli]{D.~Campana}%
\author[SSL]{W.~W.~Craig}%
\author[UH]{P.~von Doetinchem}%
\author[UCLA]{E.~Everson}%
\author[ORNL]{L.~Fabris}%
\author[UCLA]{S.~Feldman}%
\author[JAXA]{H.~Fuke}%
\author[CU]{F.~Gahbauer}
\author[UH]{C.~Gerrity}%
\author[CU]{C.~J.~Hailey}%
\author[UCLA]{T.~Hayashi}%
\author[Tokai]{A.~Kawachi}%
\author[PEDSC]{M.~Kozai}%
\author[INFN-trieste,UTrieste,IFPU]{A.~Lenni}%
\author[SSL]{A.~Lowell}%
\author[INFN-pavia,UBergamo]{M.~Manghisoni}%
\author[INFN-rome,URome]{N.~Marcelli}%
\author[SSL]{B.~Mochizuki}%
\author[PSU]{S.~A.~I.~Mognet}
\author[Shinsu]{K.~Munakata}%
\author[INFN-trieste,IFPU]{R.~Munini}%
\author[Aoyama]{Y.~Nakagami}%
\author[Heliospace]{J.~Olson}%
\author[UCLA]{R.~A.~Ong}%
\author[INFN-napoli]{G.~Osteria}%
\author[MIT,CU]{K.~M.\ Perez}
\author[UCLA]{S.~Quinn}%
\author[INFN-pavia,UBergamo]{V.~Re}%
\author[INFN-pavia,UBergamo]{E.~Riceputi}%
\author[MIT]{B.~Roach}%
\author[UCLA]{J.~Ryan}%
\author[CU]{N.~Saffold\fnref{presentFNAL}}%
\author[INFN-napoli,UNapoli]{V.~Scotti}%
\author[Kanagawa]{Y.~Shimizu}%
\author[INFN-rome,URome]{R.~Sparvoli}%
\author[UH]{A.~Stoessl}%
\author[INFN-firenze]{A.~Tiberio}%
\author[INFN-firenze]{E.~Vannuccini}%
\author[Aoyama]{T.~Wada}%
\author[MIT]{M.~Xiao}%
\author[JAXA]{M.~Yamatani}%
\author[MIT]{K.~Yee}%
\author[Aoyama]{A.~Yoshida}%
\author[JAXA]{T.~Yoshida}%
\author[INFN-trieste]{G.~Zampa}%
\author[Northeastern]{J.~Zeng}%
\author[UCLA]{J.~Zweerink}%

\address[MIT]{Massachusetts Institute of Technology, 77 Massachusetts Ave., Cambridge, MA 02139 USA.}
\address[SSL]{Space Sciences Laboratory, University of California, Berkeley, 7 Gauss Way, Berkeley, CA 94720, USA.}
\address[Northeastern]{Northeastern University, 360 Huntington Ave., Boston, MA 02115, USA.}
\address[INFN-trieste]{INFN, Sezione di Trieste, Padriciano 99, I-34149 Trieste, Italy.}
\address[IFPU]{IFPU, Via Beirut 2, I-34014 Trieste, Italy.}
\address[UCSD]{University of California, San Diego, 9500 Gilman Dr., La Jolla, CA 90037, USA.}
\address[CU]{Columbia University, 550 West 120th St., New York, NY 10027, USA.}
\address[INFN-napoli]{INFN, Sezione di Napoli, Strada Comunale Cinthia, I-80126 Naples, Italy.}
\address[UH]{University of Hawai'i at M$\overline{\textrm{a}}$noa, 2505 Correa Road, Honolulu, Hawaii 96822, USA.}
\address[UCLA]{University of California, Los Angeles, 475 Portola Plaza, Los Angeles, CA 90095, USA.}
\address[ORNL]{Oak Ridge National Laboratory, 1 Bethel Valley Rd., Oak Ridge, TN 37831, USA.}
\address[JAXA]{Institute of Space and Astronautical Science, Japan Aerospace Exploration Agency (ISAS/JAXA), Sagamihara, Kanagawa 252-5210, Japan.}
\address[Tokai]{Tokai University, Hiratsuka, Kanagawa 259-1292, Japan.}
\address[PEDSC]{Polar Environment Data Science Center,Joint Support-Center for Data Science Research,Research Organization of Information and Systems,(PEDSC, ROIS-DS), Tachikawa 190-0014, Japan.}
\address[UTrieste]{Universit\`{a} degli Studi di Trieste, Piazzale Europa 1, I-34127 Trieste, Italy.}
\address[INFN-pavia]{INFN, Sezione di Pavia, Via Agostino Bassi 6, I-27100 Pavia, Italy.}
\address[UBergamo]{Universit\`{a} di Bergamo, Viale Marconi 5, I-24044 Dalmine (BG), Italy.}
\address[INFN-rome]{INFN, Sezione di Roma ``Tor Vergata", Piazzale Aldo Moro 2, I-00133 Rome, Italy.}
\address[URome]{Universit\`{a} di Roma ``Tor Vergata", Via della Ricerca Scientifica, I-00133 Rome, Italy.}
\address[PSU]{Pennsylvania State University, 201 Old Main, University Park, PA 16802 USA.}
\address[Shinsu]{Shinshu University, Matsumoto, Nagano 390-8621, Japan.}
\address[Aoyama]{Aoyama Gakuin University, Sagamihara, Kanagawa 252-5258, Japan.}
\address[Heliospace]{Heliospace Corporation, 2448 6th St., Berkeley, CA 94710, USA.}
\address[UNapoli]{Universit\`{a} di Napoli ``Federico II", Corso Umberto I 40, I-80138 Naples, Italy.}
\address[Kanagawa]{Kanagawa University, Yokohama, Kanagawa 221-8686, Japan.}
\address[INFN-firenze]{INFN, Sezione di Firenze, via Sansone 1, I-50019 Sesto Fiorentino, Florence, Italy.}
\fntext[presentFNAL]{Present address: Fermi National Accelerator Laboratory, Wilson St.\ and Kirk Rd., Batavia IL 60510-5011, USA.}

\begin{abstract}
The General Antiparticle Spectrometer (GAPS) is an upcoming balloon mission to measure low-energy cosmic-ray antinuclei during at least three \textasciitilde35-day Antarctic flights. 
With its large geometric acceptance and novel exotic atom-based particle identification, GAPS will detect \totalcosmic~cosmic antiprotons per flight and produce a precision cosmic antiproton spectrum in the kinetic energy range of
\textasciitilde\range\,GeV/$n$ at the top of the atmosphere. 
With these high statistics extending to lower energies than any previous experiment, and with complementary sources of experimental uncertainty compared to traditional magnetic spectrometers, the GAPS antiproton measurement will be sensitive to dark matter, primordial black holes, and cosmic ray propagation. 
The antiproton measurement will also validate the GAPS antinucleus identification technique for the antideuteron and antihelium rare-event searches. 
This analysis demonstrates the GAPS sensitivity to cosmic-ray antiprotons using a full instrument simulation and event reconstruction, and including solar and atmospheric effects.
\end{abstract}


\end{frontmatter}


\section{Introduction}
\label{sec:intro}

Cosmic antinuclei present an excellent channel for detection of new physics due to their low astrophysical abundance. Antiprotons constitute <0.01\% of the cosmic particle flux at the top of Earth's atmosphere (TOA), such that precision spectral measurements are sensitive to new cosmic particle sources and to the details of Galactic propagation. Meanwhile, low-energy antideuterons and antihelium-3 nuclei are the subjects of rare event searches, and because the expected astrophysical flux is several orders of magnitude below the sensitivity of current experiments, any detection would indicate new physics. The current experimental and theoretical status for science with cosmic antinuclei is summarized in \cite{DbarReview}. Low-energy cosmic antiprotons in particular are sensitive to models of light dark matter (DM) and primordial black holes (PBHs), as well as Galactic propagation. 

Since the first detections of cosmic antiprotons in the late 1970s \cite{Golden79,Bogomolov85}, various experiments have measured the antiproton spectrum at TOA. Recent measurements by  AMS-02 \cite{AMSPbar,Aguilar21}, BESS \cite{Orito20,BessPolarPbar}, and PAMELA \cite{PamelaPbar1,PamelaPbar2} have provided information on the antiproton flux in the kinetic energy range of $0.17-450$\,GeV/$n$. {Precision antiproton measurements \cite{AMSPbar} have excluded the thermal annihilation cross section for annihilation of weakly interacting massive particles (WIMPs) with mass $M_{DM} < 40$\,GeV and $150<M_{DM} < 500$\,GeV into purely $b\bar{b}$ final states \cite{Cuoco17,Cholis19}. For $M_{DM} >200$\,GeV, antiprotons provide even more stringent limits than dwarf spheroidal galaxies \cite{Cuoco18}.}
Discussion is also ongoing in the community related to a possible excess measured by AMS-02 around $10-20$\,GeV/$n$, which has been interpreted as a signal of annihilating DM with $M_{DM}$ in the range of $40-130$\,GeV \cite{Cuoco17,Cui17, Cui18, Reinert18, Cuoco19, Cholis19}.  However, the significance of this detection depends on the treatment of systematic errors and on their correlations \cite{Cui18,Reinert18,Cuoco19,Boudaud19,Boudaud20,Heisig20,Calore22}.

{As illustrated by the robust DM limits derived from the precision antiproton spectrum at higher energies,} a precision low-energy (<$0.25$\,GeV/$n$) cosmic antiproton spectrum would open sensitivity to possible new physics. 
In particular, several hidden sector DM models naturally predict large cosmic particle signals at low energies \cite{HooperHiddenSector}. 
{Such a measurement could also constrain the low-energy edge of the possible DM signal that would explain the reported  $10-20$\,GeV excess.}
Beyond DM, evaporation of local, light (formed with mass M$_{\star} \sim 5 \times 10^{14}$\,g; not responsible for the cosmological DM abundance \cite{Page76,Villanueva21}) PBH could produce an excess of antiprotons broadly peaked around 0.5\,GeV \cite{HawkingRadiation, HawkingPBH,Carr2020PBH,PBHPbar,Herms17}. 
Because this source falls off less steeply at low energies than the expected secondary astrophysical flux of sub-GeV antiprotons, a PBH excess could appear below 0.25\,GeV/$n$ despite the lack of detection at higher energies by the BESS, PAMELA, and AMS-02 programs.

{Interpretation of cosmic particle flux measurements relies on a precisely tuned model of particle transport in the Galaxy and in the heliosphere, with its attendant uncertainties due to both tuning of propagation parameters and knowledge of interaction cross sections. 
Low-energy antiprotons are sensitive to the Galactic and solar conditions affecting cosmic particle transport because 1) cosmic antiprotons arise principally from spallation of cosmic-ray nuclei on the interstellar medium and 2) low-energy particles are strongly deflected by magnetic fields \cite{Strong07}. 
A high-statistics low-energy antiproton spectrum will force a comparison with Galactic and solar propagation models that have been tuned to measurements at higher energies.}

The General Antiparticle Spectrometer (GAPS)  \cite{Mori,HaileyGAPS1,HaileyGAPS2,AntideuteronAramakiSensitivity,AntiprotonAramakiSensitivity,GapsHeBarSensitivity} is a NASA Antarctic long-duration balloon (LDB) mission designed specifically for detection of low-energy cosmic antinuclei. 
With its novel particle identification method based on exotic atom formation and decay, and the low geomagnetic cutoff expected from an Antarctic flight path, GAPS is poised to extend the precision cosmic antiproton spectrum to a new low-energy range. Notably, the GAPS detection method features {both larger acceptance in a lower-energy range and} complementary sources of instrument systematic uncertainty compared to traditional magnetic spectrometer techniques.

The potential of GAPS to provide unprecedented sensitivity to low-energy antiprotons \cite{AntideuteronAramakiSensitivity} 
has previously been established using preliminary simulation and analysis tools\cite{AntiprotonAramakiSensitivity}. 
This work uses a detailed detector simulation with full event reconstruction to assess the sensitivity of the GAPS instrument to cosmic antiprotons in its first flight and through its full program of three or more LDB flights. Sec.\ \ref{sec:gaps} introduces the GAPS experimental design and detection scheme. The instrument simulation framework is discussed in Sec.\ \ref{sec:sims}, and fluxes are calculated using a separate simulation described in Sec.\ \ref{sec:fluxes}. Sec.\ \ref{sec:analysis} details the analysis strategy for antiproton identification against abundant backgrounds. Finally, Sec.\ \ref{sec:conclusions} summarizes the outlook for the GAPS antiproton measurement.

\begin{figure*}[tbh]
\centering
\includegraphics[width=0.48\textwidth,trim = 180 0 150 0,clip]{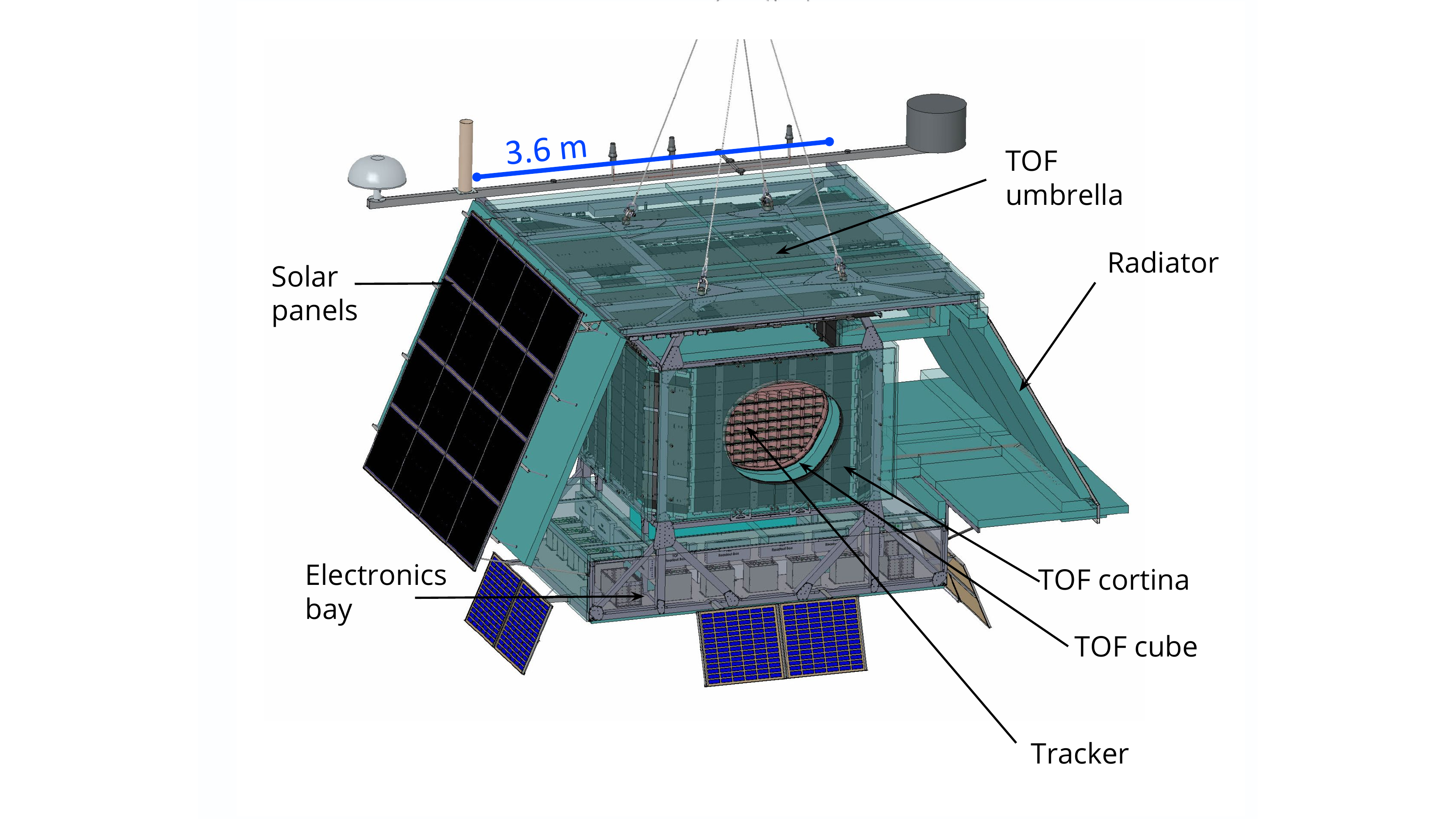}
\includegraphics[angle=90,width=0.48\textwidth, trim= 100 170 0 100, clip]{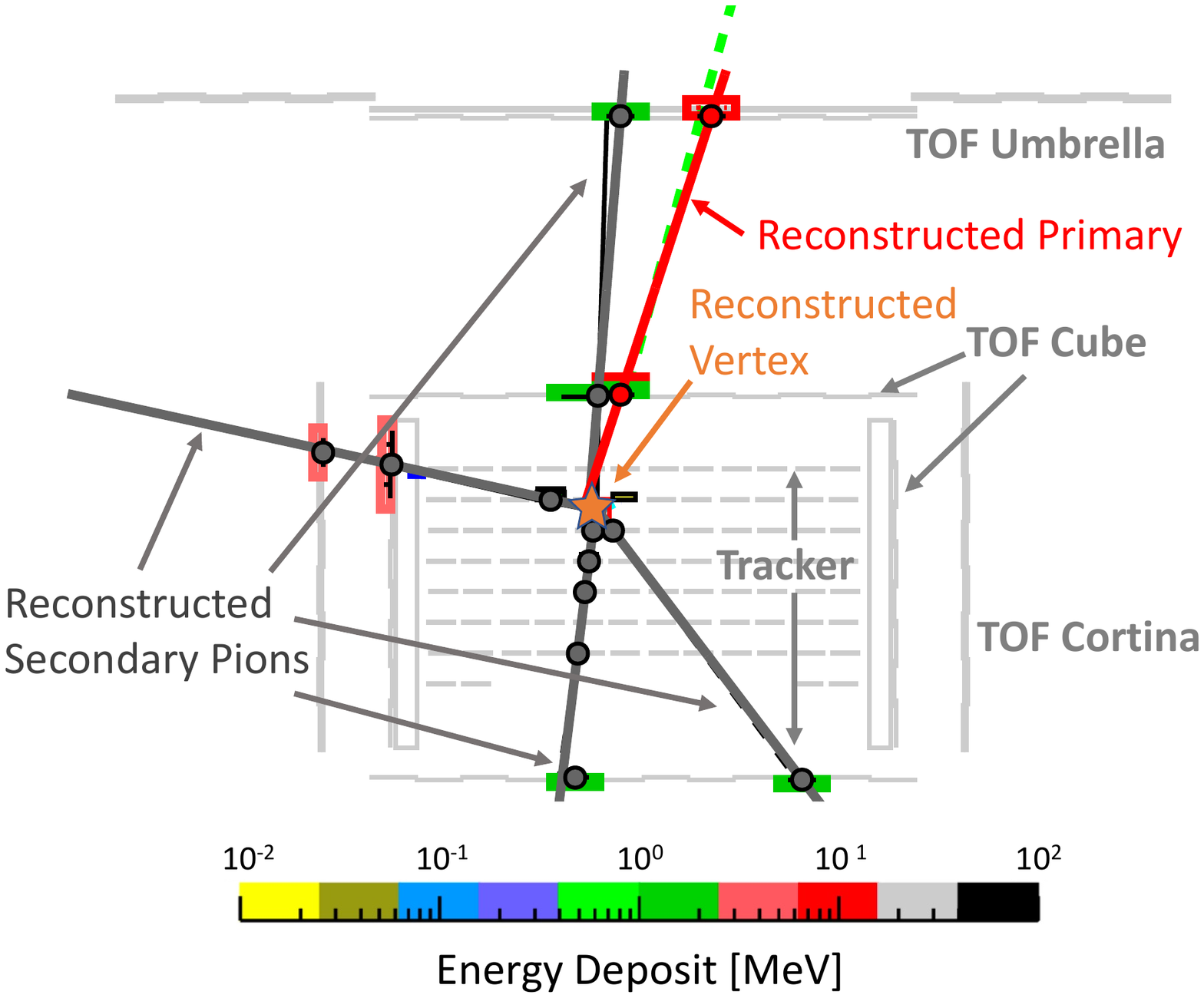}
\caption{\label{fig:gaps-inst}
{\em Left}: A mechanical drawing of the GAPS instrument illustrates the outer TOF umbrella and cortina and the inner TOF cube. The cut-away panels reveal the layers of detectors in the Si(Li) tracker. The electronics bay is located beneath the sensitive material while the solar panels and the radiator for the oscillating heat pipe thermal system are to the side, minimizing the mass directly above the science payload. {\em Right}: A reconstructed signal event (simulated with $\beta = 0.37$) shows the primary antiproton track (simulated green, reconstructed red), four secondary pion tracks (simulated black, reconstructed dark gray), and the annihilation vertex (reconstructed orange star). The boxes highlight sensitive detector volumes in which energy was deposited, where the color bar gives the total energy deposition in MeV. The largest energy depositions (red) are on the primary track. The remaining sensitive detector materials are represented in light gray. 
}
\end{figure*}


\section{The GAPS Experiment}\label{sec:gaps}

\subsection{The GAPS Instrument}\label{sec:gaps_det}

Fig.~\ref{fig:gaps-inst} ($left$) illustrates the GAPS instrument, which consists of a 
silicon tracker and X-ray spectrometer system (hereafter the ``tracker'') surrounded by a plastic scintillator time-of-flight (TOF) system. The electronics bay is located below the sensitive volume and the solar panels and radiator are located to the side to minimize material between the atmosphere and the sensitive detectors.

The GAPS tracker is designed to accommodate 1440 10\,cm-diameter, 2.5\,mm-thick lithium-drifted silicon (Si(Li)) detectors \cite{PerezSiLi, RogersSiLi, KozaiSiLi, RogersIEEE, KozaiIEEE, Kozai21,Semikon} arranged in 10 planes to fill a $1.6 \mathrm{\,m} \times 1.6 \mathrm{\,m} \times 1 \mathrm{\,m}$ volume. 
The detectors are mounted in aluminum modules of four detectors each, with front-end readout electronics arranged between the detectors, and the layers are separated by polyethylene foam. 
The Si(Li) system serves as the target material, particle tracker, and X-ray spectrometer for particle identification. 
Each 8-strip Si(Li) detector provides $\lesssim$4\,keV energy resolution for X-rays in the $20 - 100$\,keV range and <10\% resolution up to 100\,MeV. The dynamic range is enabled by the novel signal compression technique used for readout by the custom ASIC \cite{Manghisoni15,Manghisoni18,ScottiICRC,Manghisoni21}.
The detectors are passivated for long-term stability \cite{SaffoldSiLi} and are operable in ambient pressure both on the ground and in flight. 
They function at temperatures in the range of $-50\degree$C to $-30\degree\textrm{C}$, achievable in flight using an integrated oscillating heat pipe (OHP) thermal system \cite{CapillaryHeatPipe,OHP,FukeOHPIEEE} together with passive radiation. The ability to fly without a massive pressure vessel or cryostat facilitates the large sensitive volume within the mass, volume, and power constraints of an LDB payload. 
For the first flight, the tracker will be instrumented with approximately 1000 detectors, with the remaining spaces filled with aluminum blanks. 

The GAPS TOF \cite{QuinnICRC19,QuinnICRC21} is composed of 160 $16\,\mathrm{cm} \times 0.63\,\mathrm{cm} \times 108-180\,\mathrm{cm }$ plastic scintillator paddles, each read out by six silicon photomultipliers at either end. 
The inner TOF ``cube'' structure completely encloses the tracker. 
Meanwhile, the outer TOF consists of two distinct structures: the ``umbrella,'' a plane of paddles \textasciitilde90\,cm above the cube, and the ``cortina,'' which surrounds the sides of the cube at a distance of 30\,cm. 
Together, the umbrella, cortina, and cube ensure that at least two hits are recorded for downward-going particles over a wide solid angle. 
The paddles provide timing resolution of <400\,ps and spatial resolution of \textasciitilde4\,cm in the lateral direction, for a typical $\beta$ resolution of {$0.015 - 0.02$} (RMS) {for cosmic particles in the GAPS energy range that traverse the umbrella}. The TOF provides the GAPS system trigger using logic based on energy deposition and number of hits. Structural support is provided by an {aluminum} frame.

In 2012, a prototype GAPS flight \cite{DoetinchemPGAPS,MognetPGAPS,FukePGAPS} demonstrated operation of the tracker and TOF detector components at high altitude{, with a piggy-back test of the OHP} \cite{Fuke17}. Integration of the hardware and software systems for the first GAPS flight is underway.

\subsection{The GAPS Particle Identification Concept}\label{sec:gaps_detection}
The novel GAPS particle identification concept is based on the formation, de-excitation, and annihilation of exotic atoms. The products of exotic atom de-excitation and annihilation, together with ionization loss patterns on the primary track, are the basis for identifying rare antideuterons and antihelium-3 nuclei from a background of relatively abundant antiprotons. 
Meanwhile, this study focuses on the identification of antiprotons from a background of protons and heavier nuclei based on the existence of an annihilation signature. 

Fig.~\ref{fig:gaps-inst} ($right$) displays a simulated antiproton event. When a low-energy antinucleus traverses the GAPS instrument, it first crosses the outer (umbrella or cortina) and inner (cube) TOF layers, which measure the kinematic variable $\beta=v/c$, where $v$ is the particle velocity and $c$ is the speed of light. It then slows down via ionization and excitation losses in the tracker layers, with energy depositions characteristic of its charge and energy. 
Once the kinetic energy of the antinucleus is similar to the binding energy of the nearby target atoms (typically Al or Si), it is captured by a target atom, forming an exotic atom in an excited state. 
Within $\mathcal{O}(1)$\,ns of formation, the exotic atom de-excites via emission of Auger electrons and X-rays. The X-rays can be detected in the surrounding Si(Li) detectors, with energies uniquely determined by the mass and charge of the antinucleus and target atom \cite{KEKBeam}. 
The captured antinucleus then annihilates with the target nucleus, producing secondary hadrons. 
The characteristic ``annihilation star'' signature of exotic atom formation and decay consists of secondary tracks emerging from the annihilation vertex. 

For a given $\beta$, the different antinucleus species are distinguished based on the energy deposited on the primary track, the depth through which the energy was deposited, the de-excitation X-rays, and the number of secondary tracks. 

Positive nuclei do not form exotic atoms, and in the GAPS energy range they typically stop in the tracker without producing secondary tracks, as distinct from antinuclei. 
However, with increasing $\beta$, the cross section for hard interactions with target nuclei increases relative to the cross section for ionization losses. 
For $\beta \gtrsim 0.4$, hard interactions of both primary positive nuclei and antinuclei are increasingly common. 
Because protons and heavier positive nuclei outnumber antiprotons by a factor of $>$$10^6$ in this energy range, even rare hard interactions of primary positive nuclei, whose interaction products can mimic the antinucleus event signature, present an important background to the antiproton measurement.  {Positive nuclei arriving with $\beta > 0.7$, outside of the GAPS energy range, also present a key background if they either 1) are wrongly reconstructed with $\beta$ in the GAPS range before undergoing hard interactions or 2) interact in the TOF to produce a slow antiproton which is reconstructed.}

Antinucleus identification amidst the cosmic-ray particle background, which forms the basis of the antiproton measurement, also critically paves the way for rare event searches amongst the antinucleus events.  
While X-rays are not treated in this study and are not required for nucleus-antinucleus discrimination, they will enhance signal identification power for the antideuteron and antihelium-3 nuclei searches. 
 An antiproton analysis requiring identification of antiprotonic X-rays is deferred to a future publication; such an analysis will validate the X-ray signature for the rare event searches.

\section{Detector Simulation and Event Reconstruction}\label{sec:sims}

This analysis makes use of a detector simulation based on the {\tt Geant4} \cite{Agostinelli03,Allison06,Allison16} framework to model interactions of cosmic-ray particles with the GAPS instrument. 
The GAPS simulation includes a detailed implementation of the Si(Li) detectors and their associated passive readout and support material in the tracker, and of the scintillator paddles and associated passive material in the TOF. 
The simulation used in this analysis assumes 1000 Si(Li) detectors in the tracker, as in the first GAPS flight. It reflects the final flight geometry save for a slight modification of the corner paddles of the TOF cortina, whose mechanical implementation had not been finalized.

Particles are generated isotropically over the 2$\pi$ solid angle of the downward momentum direction from the surface of a 4.4\,m cube encapsulating the GAPS instrument. 
The generated $\beta$ is uniform in the range of $0.1\leq\beta<0.7$ and uniform with higher statistics in the range of $0.7 \leq \beta < 1.0$\textcolor{red}{}. 
The physics list {\tt FTFP\_BERT\_HP} is used in {\tt Geant4} v10.07 to simulate interactions of these particles in the sensitive and passive detector materials. This analysis uses $1.2\times10^9$ simulated antiprotons, $2.8\times10^{12}$ simulated protons, and $4.8\times10^{11}$ simulated $^4$He nuclei. Of these, $3.5\times10^7$  antiprotons, $3.1\times10^8$ protons, and $3.6\times10^8$ $^4$He nuclei pass the trigger conditions (Sec.~\ref{sec:trig}) for use in the analysis. 
{Less abundant nuclei are not simulated for this analysis. As demonstrated in Ref.~\cite{GapsHeBarSensitivity,StoesslICRC21}, carbon and higher-$Z$ nuclei are effectively rejected by the charge selection described in Sec.~\ref{sec:presel}. 
Deuteron and $^3$He nucleus fluxes are subdominant to those of protons and $^4$He nuclei.
}

Fig.~\ref{fig:gaps-inst} ($right$) illustrates a reconstructed antiproton event.
For every generated event, each energy deposition (hit) in an active detector (TOF paddle or Si(Li) strip) is recorded and convolved with the timing, position, and energy resolution of the respective detector element.
Events passing the antiparticle trigger conditions (Sec.\ \ref{sec:trig}) are passed through a reconstruction sequence designed specifically for the GAPS annihilation star event topology \cite{Reconstruction}. 
First, the primary track is reconstructed. The primary-track hits in the outer and inner TOF are identified based on their timestamps and energy depositions, while hits in the Si(Li) tracker are subsequently associated with the primary track based on the compatibility of their positions and energy depositions with the initial TOF hits. 
For \textasciitilde85\% of {antinuclei interacting} within the {volume defined by the tracker system}, at most one hit is either wrongly associated with or wrongly missing from the reconstructed primary. 
The reconstructed primary trajectory is based on a least-squares fit to all of the hits on the primary track, and the primary $\beta$ is calculated using the TOF timing information with this trajectory. 
A custom algorithm identifies secondary tracks based on the remaining TOF and Si(Li) hits not associated with the primary track. 
For events with two or more hits on the primary track and with at least one secondary track with two or more hits, the {interaction} vertex position is identified by minimizing the distance of closest approach to each secondary track. 
The efficiency for reconstructing an {interaction} vertex is \textasciitilde90\%, with a most probable {absolute displacement} of 9\,mm  from the true vertex and 68\% of events reconstructed within 104\,mm of the true vertex \cite{Reconstruction}. 

Following reconstruction, a correction is applied to the primary $\beta$ to account for the typical energy loss by $|Z| = 1$ particles in the outer TOF paddle. More sophisticated correction techniques, which will result in improved resolution for the reconstructed $\beta$, are under development.

\begin{figure*}[tbh]
\centering
\includegraphics[width = 0.48\textwidth]{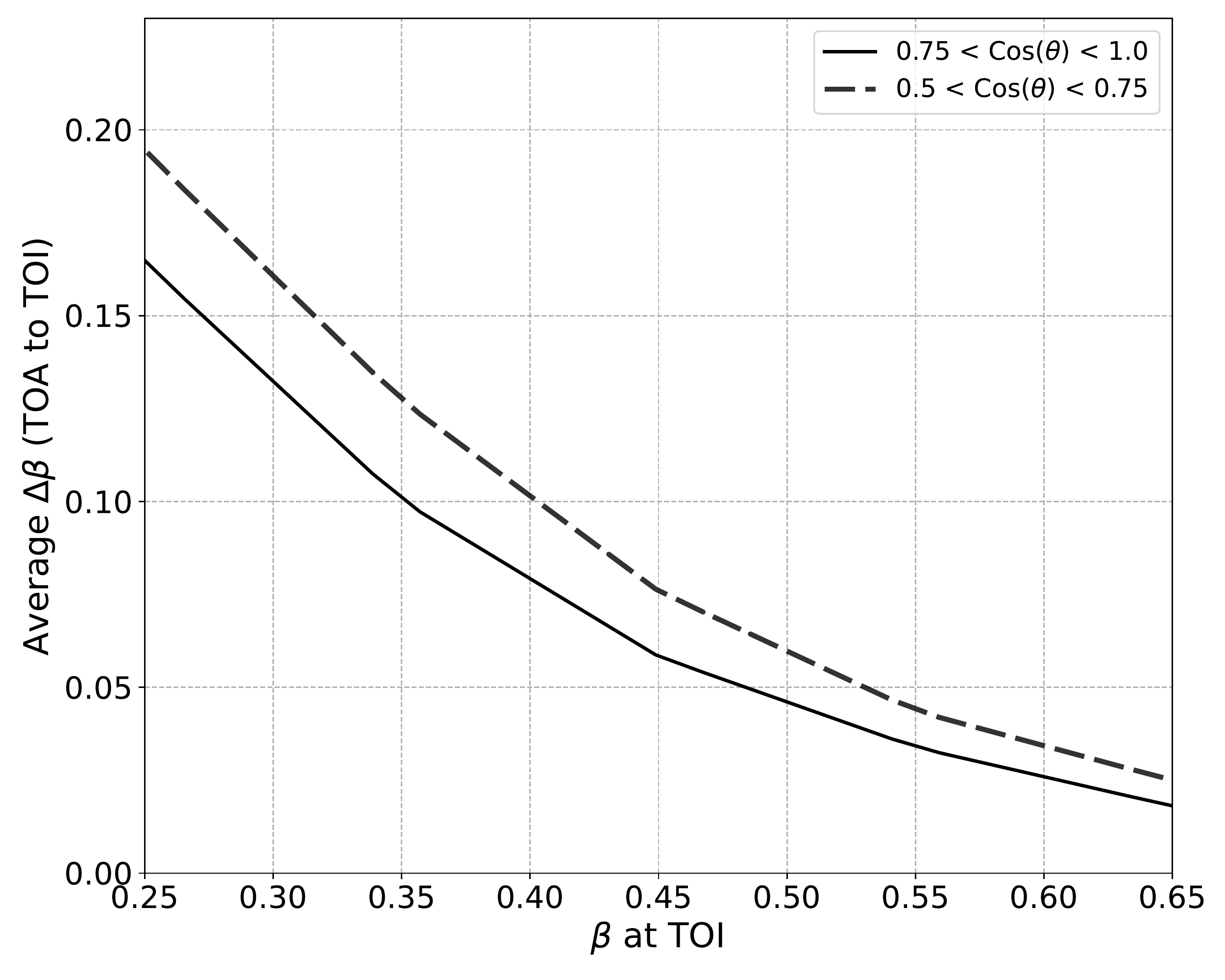}
\includegraphics[width = 0.48\textwidth]{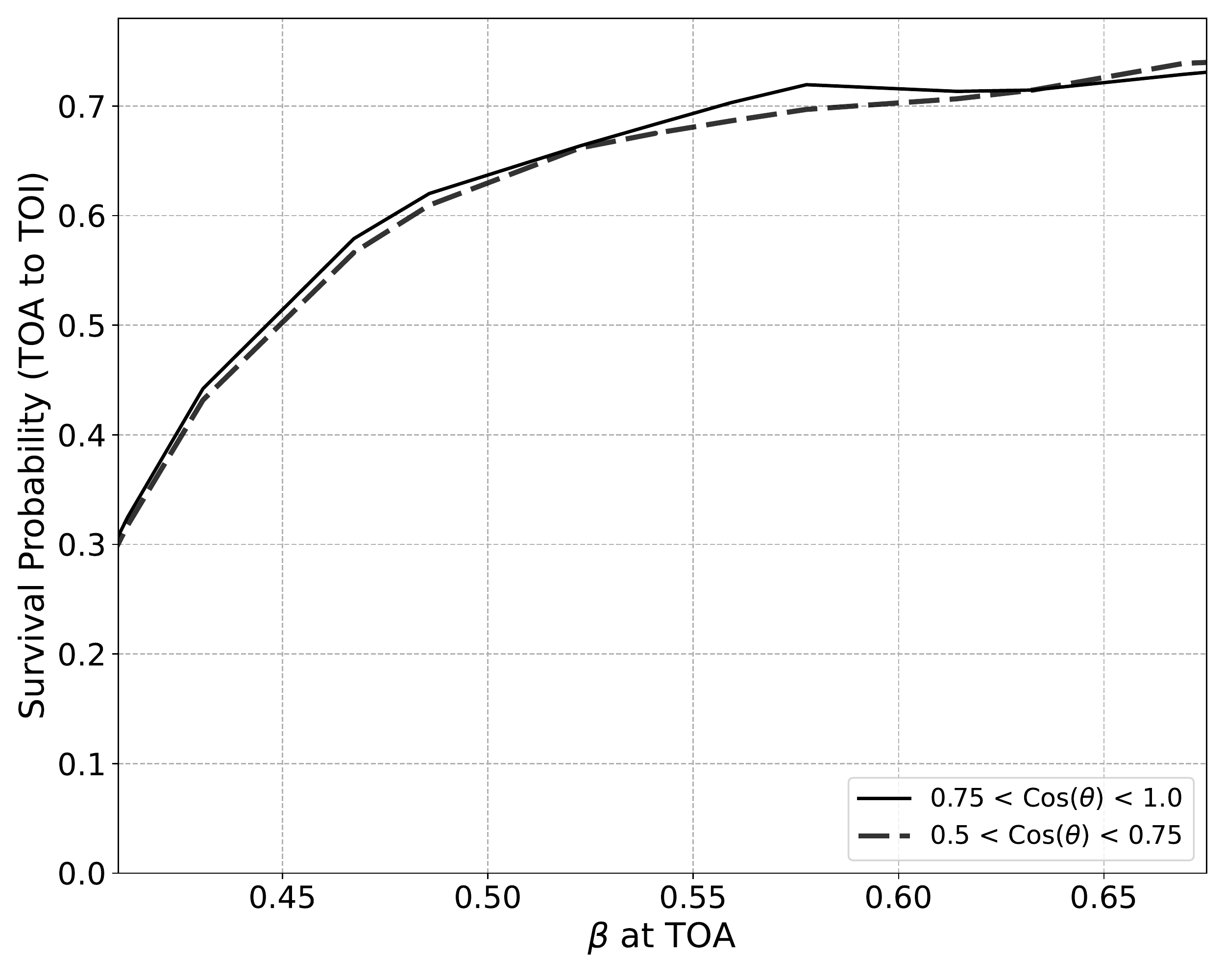}
\caption{\label{fig:atmos} $Left$: The average velocity loss for antiprotons from the top of the atmosphere (TOA) to the top of instrument (TOI) is shown as a function of $\beta$ at TOI. The decrease depends on the zenith angle $\theta$, defined such that $\cos\theta = 1$ indicates a vertical trajectory, and is presented here in two bins relevant for the cosmic antiproton analysis: $0.75 < \cos\theta < 1.0$ (solid) and $0.5 < \cos\theta < 0.75$ (dash). The sensitive range of {$0.25\lesssim\beta\lesssim0.65$} at TOI corresponds to {$0.41\lesssim\beta\lesssim0.68$} at TOA. 
$Right$: The survival probability for antiprotons at TOA to reach TOI is given as a function of $\beta$ at TOA and presented in the same angular bins.
}
\end{figure*}

\begin{figure*}[hbth]
\centering
\includegraphics[width = 0.48\textwidth]{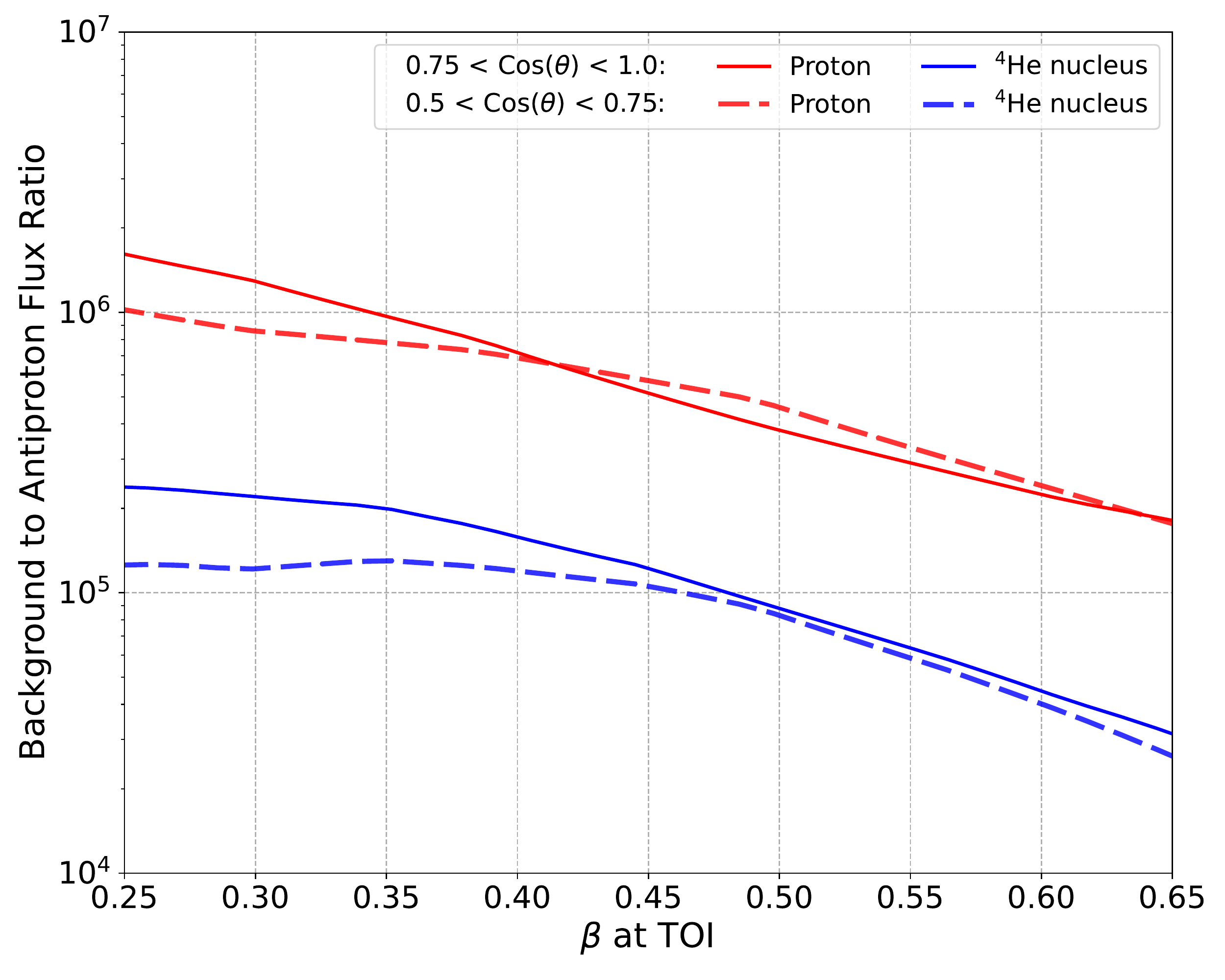}
\includegraphics[width = 0.48\textwidth]{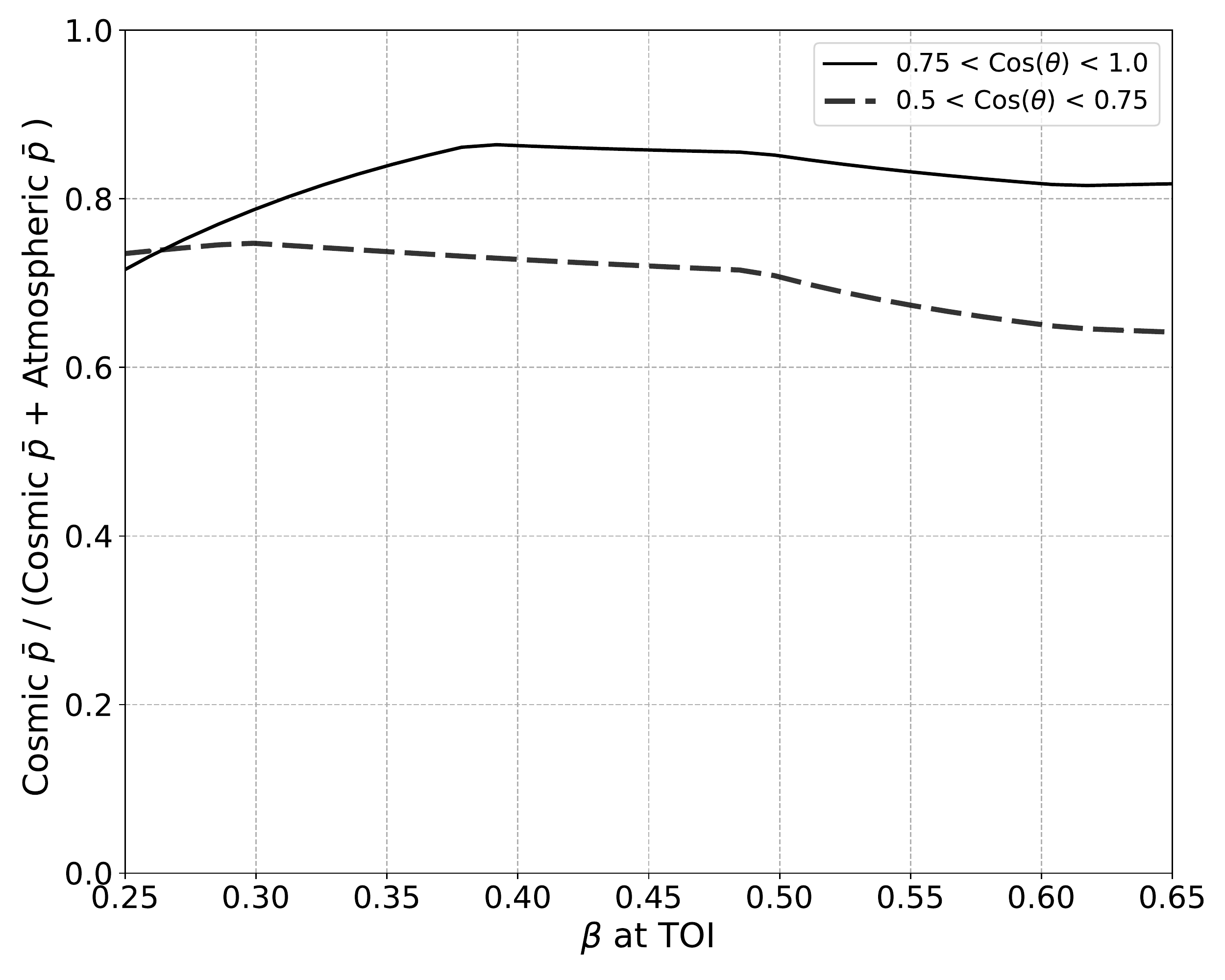}
\caption{\label{fig:fluxes} $Left:$ The ratios of the proton (red) and $^4$He nucleus (blue) background fluxes to the total (cosmic and atmospheric) antiproton flux are shown as a function of $\beta$ at TOI. The ratios for each species are binned by the zenith angle $\theta$, where $\cos\theta = 1$ indicates a vertical trajectory, and are shown for $0.75 < \cos \theta < 1.0$ (solid) and $0.5 < \cos \theta < 0.75$ (dash). 
$Right$: The ratio of the cosmic antiproton flux to the total antiproton flux at TOI is shown as a function of $\beta$ in the same angular bins. 
}
\end{figure*}


\section{Simulation of Particle Fluxes}\label{sec:fluxes} 

Particle fluxes at float altitude (top of instrument; TOI) are calculated by modulating the local interstellar spectra with solar, geomagnetic, and atmospheric effects. 
Fluxes for cosmic antiprotons, as well as isotopes of hydrogen, helium, and heavier positive nuclei in the local interstellar region are simulated using {\tt Galprop} \cite{Strong98,Strong07}, with propagation tuned to match {\em PAMELA} and {\em Voyager I} data as in \cite{Bisschoff19}. 
These local interstellar spectra are modulated according to the solar activity anticipated for the Austral summer of $2022-23$, following the model in \cite{SolarModModel,MuniniPosMod} to produce the flux at TOA. 

{A model of particle energy loss and absorption in the atmosphere is required to predict fluxes for all particle species at TOI and to transform any GAPS measurement into a TOA measurement. }
Atmospheric effects are calculated assuming a 37\,km float altitude using the {\tt PLANETOCOSMICS} \cite{Planetocosmics,PVDThesis} simulation package updated to run with {\tt Geant4} v10.06. 
Assuming a realistic LDB flight trajectory uniformly distributed from $-78\degree$ to $-85\degree$ latitude, geomagnetic modulation allows $60-80\%$ survival in the GAPS energy range, where survival increases with particle rigidity. 

Fig.~\ref{fig:atmos} illustrates the effects of atmospheric and geomagnetic attenuation on cosmic antiproton fluxes. 
The atmosphere introduces an angular dependence in the fluxes at float altitude, as the amount of atmosphere traversed varies with the zenith angle $\theta$, where $\cos\theta=1$ indicates a vertically downward trajectory. 
The average velocity attenuation due to ionization and excitation losses in the atmosphere depends on $\beta$ as well as $\theta$. 
Cosmic particle fluxes also decrease due to absorption in the atmosphere as well as geomagnetic modulation, such that a particle at TOA has a $\beta$- and $\theta$-dependent probability of surviving to TOI. Antiprotons with $\beta \lesssim 0.4$ at TOA are strongly absorbed and are unlikely to reach TOI.

Inelastic collisions of energetic particles in the atmosphere lead to the production of secondary ``atmospheric'' particle fluxes at TOI. Thus, the total (cosmic and atmospheric) particle fluxes observed by GAPS depend on the full cosmic-ray energy spectrum at TOA, and the atmospheric component varies with the depth of the atmosphere and thus with $\theta$. 
For antiprotons in particular, most of the cosmic flux is produced in collisions of cosmic-ray protons and helium nuclei with the interstellar medium. 
Because the grammage traversed by particles from the top of the atmosphere to float altitude is comparable to the grammage traversed by a typical particle in its journey through the Galaxy, the flux of atmospheric antiprotons is comparable to that of cosmic antiprotons at float altitude.
{Particles arriving at wider angles have traversed more atmosphere, with correspondingly increased velocity attenuation and opportunity for production of atmospheric antiprotons.}

Fig.~\ref{fig:fluxes} illustrates the background fluxes for the GAPS antiproton measurement. 
Protons and $^4$He nuclei are the most abundant background species, with the proton-to-antiproton flux ratio exceeding $10^6$ in the lower-$\beta$ portion of the GAPS energy range. {The flux of background particles relative to antiprotons drives the background rejection power required in the analysis.} 
Meanwhile, atmospheric antiprotons present an irreducible background for the cosmic antiproton measurement. 
Cosmic antiprotons dominate the total flux for $\cos\theta > 0.5$, accounting for >70\% of the total flux at the peak of the antiproton acceptance ($\beta\sim0.4$). In contrast, for $\cos\theta < 0.5$, atmospheric antiprotons dominate the total flux, inherently limiting the possible precision of a cosmic antiproton spectrum. Thus, only antiprotons arriving with $\cos\theta > 0.5$ are treated in this analysis. Antiprotons with $\cos\theta < 0.5$ will be used to tune the atmospheric model, necessary to control systematic errors for all GAPS measurements. 
{The GAPS sensitivity to antiprotons arriving with $\cos\theta < 0.5$ will be treated in a future publication dedicated to atmospheric fluxes.}

\section{GAPS Antiproton Analysis}\label{sec:analysis}

The figure of merit for sensitivity to cosmic particles is the acceptance for signal and background species. 
The acceptance $\Gamma_a (E)\ [\textrm{m}^{2}\textrm{sr}]$ describes the physical extent of the instrument modified by the $\beta$-dependent efficiency for particle species $a$ to pass any selection criteria. 
Given $\Gamma_a (E)$, the expected number d$N_a(E)/$d$E$ of particles of species $a$ per unit energy is calculated via:
\begin{equation}\label{eq:accep}
\textrm{d}N_a(E)/\textrm{d}E =  \Phi_a (E) \cdot  \Gamma_a 
(E)  \cdot  T 
.
\end{equation}
Equation \eqref{eq:accep} shows that
d$N_a(E)/$d$E$ depends on the flux $\Phi_a(E)\ [\textrm{s}^{-1}\textrm{m}^{-2}\textrm{sr}^{-1}\textrm{(GeV/$n$)}^{-1}]$ and instrument livetime $T\ [\textrm{s}]$ as well as $\Gamma_a(E)$.

This section treats the calculation of the $\Gamma_a(E)$ factor, which depends on the instrument geometry and particle identification, {in two angular regions}. The fluxes $\Phi_a(E)$ are calculated as described in Sec.~\ref{sec:fluxes} {for the same angular regions}. 
The multi-step analysis to identify antiproton events consists of trigger conditions in Sec.~\ref{sec:trig}, event-quality selection criteria in Sec.~\ref{sec:presel},  
and finally a likelihood analysis in Sec.~\ref{sec:llh}. The final acceptance is calculated in Sec.~\ref{sec:accep} and the resulting statistics are presented in spectral form in Sec.~\ref{sec:toi}.

\subsection{Trigger Conditions}\label{sec:trig}
The analysis is based on trigger conditions designed to select antinuclei during the GAPS flight. 
The trigger requires at least eight hits in the TOF, distributed with at least three each in the outer and inner TOF. 
This requirement selects events that interact in the instrument to produce secondary tracks. 
The trigger also requires the two largest TOF energy depositions 
to be consistent with 
$|Z| = 1$ or $|Z| = 2$ particles with $ 0.2 <\beta< 0.6$. 
This rejects highly relativistic protons based on their low ionization losses as well as primaries with $|Z| \geq 6$ based on their high ionization losses. 
The trigger provides a rejection factor of approximately 700 (50) for protons ($^4$He nuclei) while retaining >60\% of antinuclei in the GAPS $\beta$ range \cite{QuinnICRC19}.

\subsection{Event Preselection}\label{sec:presel} 
Events are selected that have primary $\beta$ reconstructed in the range of $0.25 < \beta < 0.65$. 
Only events reconstructed with $\cos\theta>0.5$, where cosmic antiprotons dominate the total flux, are selected. Additionally, this analysis only uses events in which the reconstructed primary traverses the TOF umbrella and cube, and thus where the $\beta$ resolution $\Delta\beta \lesssim 0.02$, suitable for a precision spectrum. 

{To ensure a sample of events with well-reconstructed topology, events} are rejected if the reconstruction algorithm does not converge. {As discussed in Sec.~\ref{sec:sims}, this affects <10\% of triggered antiprotons with $0.25<\beta<0.65$ \cite{Reconstruction}.} Events are also rejected if >1 sensitive detector intersecting the path of the reconstructed primary is without a hit or if the reconstructed vertex is outside the volume enclosed by the TOF cube.

Only events with ionization losses consistent with charge $|Z| = 1$ primaries with $0.25<\beta<0.65$ are used in the analysis. 
In the GAPS energy range, the typical energy loss per distance traveled (d$E$/d$x$) is proportional to $Z^2/\beta^2$ in a given material. 
The {\bf primary truncated mean d{\em E}/d{\em x}} variable defined in Sec.~\ref{sec:llh} characterizes the initial \dedx of the primary track. 
Distributions of this variable are constructed using simulated antiprotons as a function of $\beta$, and events are required to fall within the central 90\% of the distribution to be included in the analysis, where the high and low thresholds are functions of $\beta$.
This selection criterion (cut) rejects particles with $|Z|\geq3 $. Over 99\% of $^4$He nuclei reconstructed within $\Delta\beta < 0.06$ of their true $\beta$ are also rejected, though more higher-$\beta$ $^4$He nuclei persist.

\begin{figure*}[bth]
\centering
\includegraphics[width=0.48\textwidth, trim={0 0 0 0}]{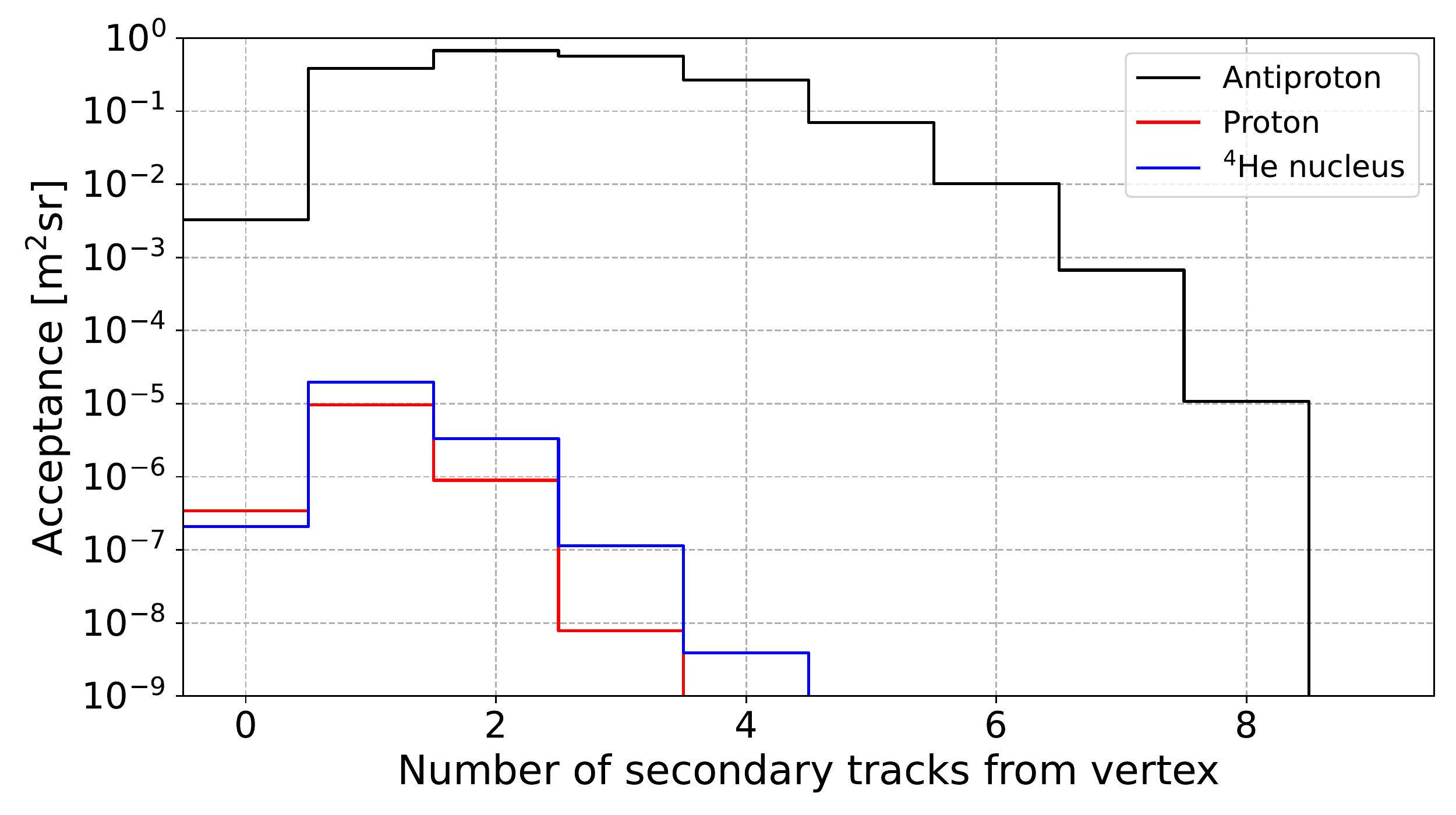}
\includegraphics[width=0.48\textwidth, trim={0 0 0 0}]{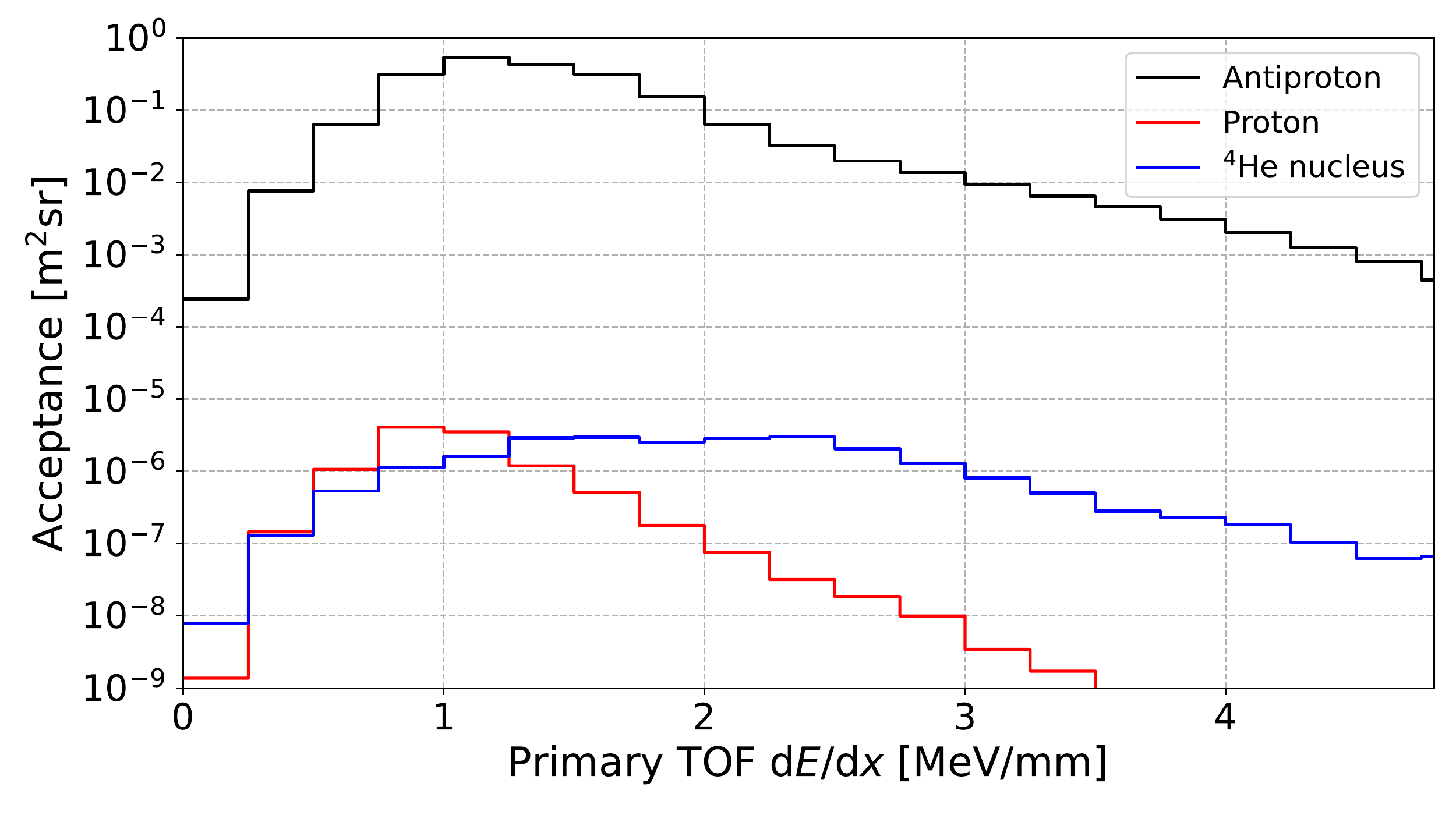}
\includegraphics[width=0.48\textwidth, trim={0 0 0 0}]{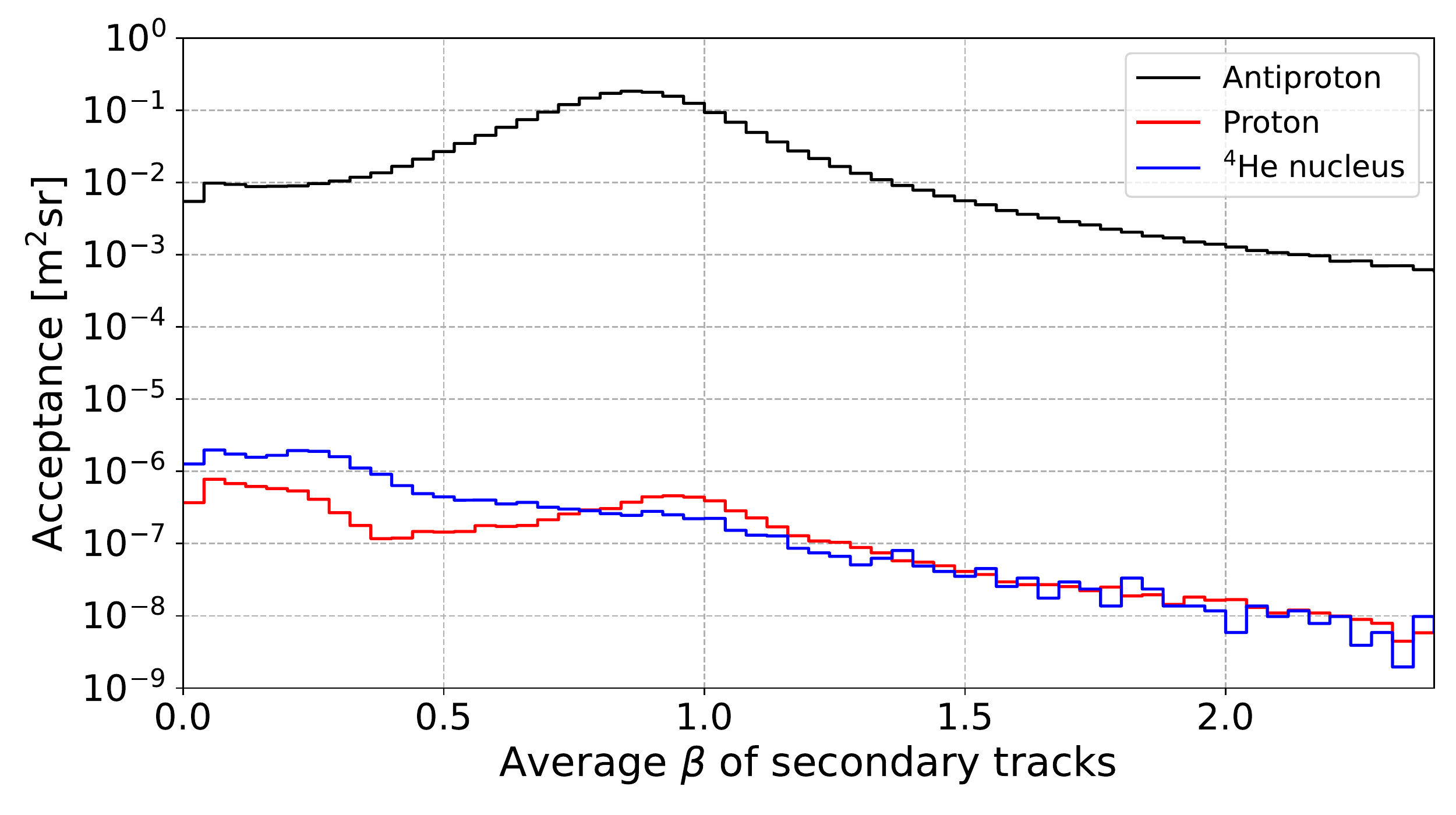}
\includegraphics[width=0.48\textwidth, trim={0 0 0 0}]{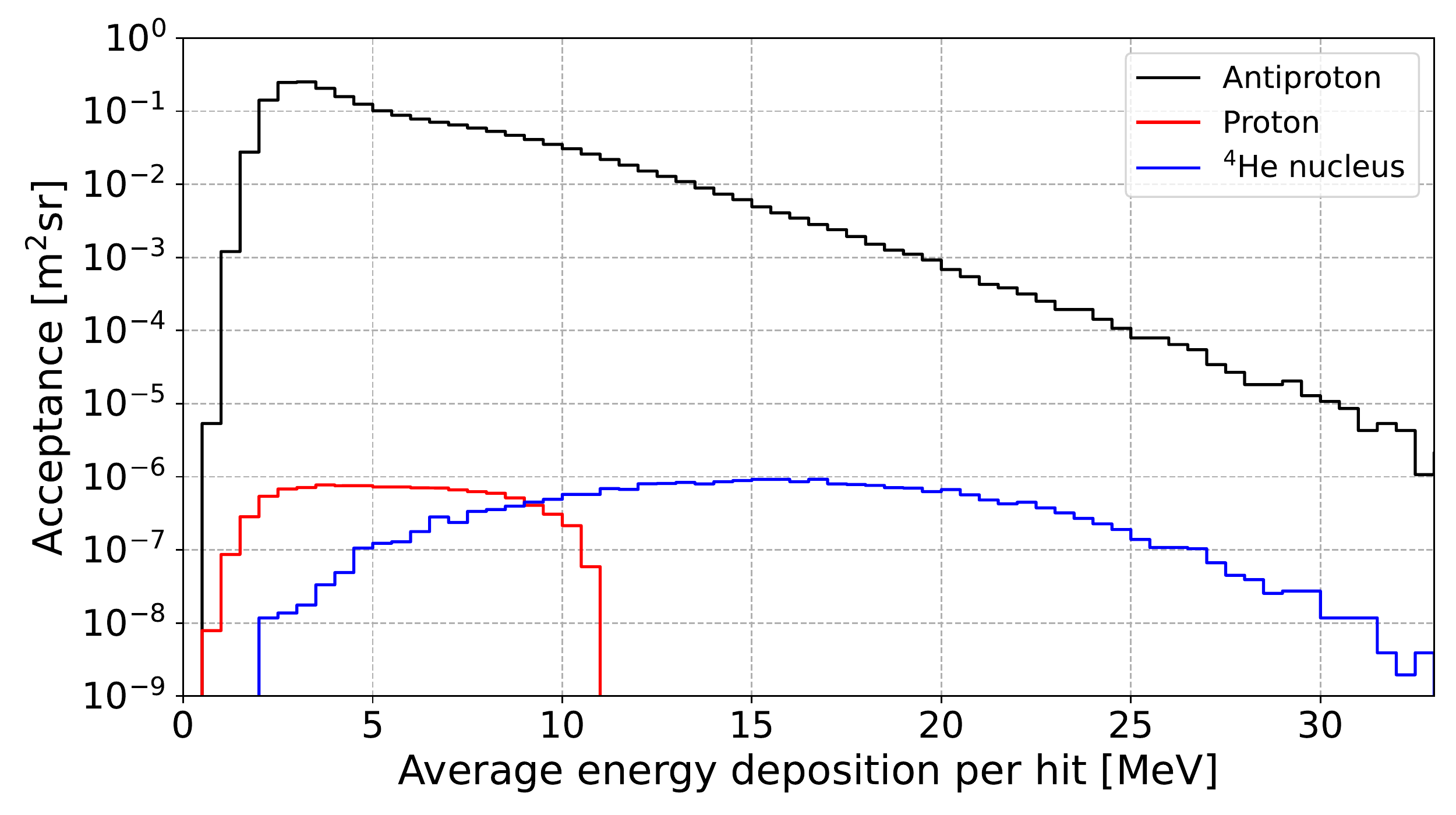}
\caption{\label{fig:vars}
Distributions of four event variables are shown for triggered and reconstructed antiprotons (black), protons (red), and $^4$He nuclei (blue) with true velocity in the range of $0.35<\beta<0.45$. 
{\em Upper left}: Number of secondary tracks from vertex. {\em Upper right}: Primary TOF d{\em E}/d{\em x}. {\em Lower left}: Average $\beta$ of secondary tracks. {\em Lower right}: Average energy deposition per hit.}
\end{figure*}

\subsection{Particle Identification}\label{sec:llh}
This section describes the particle identification tools developed to reject positive nucleus background events that pass trigger conditions due to production of secondary particles through hard interactions in the detector. 
Such events represent a small fraction of the total positive nucleus flux incident on GAPS. {However, due to the signal-to-background flux ratios,  positive nuclei still} outnumber the antiprotons passing trigger and preselection conditions, requiring a robust analysis to produce a clean sample of antiprotons. 

Particle identification is based on two likelihood classifiers. The ``$\beta$-reconstruction likelihood'' is constructed to target high-$\beta$ background events that are incorrectly reconstructed in the GAPS $\beta$ range. As discussed in Sec.~\ref{sec:gaps_detection}, high-$\beta$ background events can appear in the GAPS $\beta$ region due to 1) $\beta$ resolution effects or 2) hard interactions in an outer TOF paddle resulting in the production of an `instrumental' antiproton that is subsequently reconstructed. 
In complement, the ``identification likelihood'' is targeted toward background events with good $\beta$ reconstruction. 

Particle identification is enabled by differences in the signal and background event topologies. 
Hard interactions of background nuclei with target nuclei produce secondary particles that, due to baryon number conservation, are typically slower and less numerous compared to those arising from antiproton-nucleus annihilation. Additionally, these hard interactions always occur in flight while antiproton-nucleus annihilations can occur in flight or at rest following formation of an exotic atom. Compared to annihilation at rest, hard interactions lack the distinct $\beta$ dependence of the
ionization loss pattern on the primary track, and the interactions are boosted in the direction of the primary momentum. 
In addition, high-$\beta$ background events with $Z = 1$ are rejected based on their overall lower ionization losses compared to antiprotons in the GAPS energy range. Instrumental antiprotons are rejected based on the anomalously large energy deposition in the outer TOF corresponding to the hard interaction site. 

The two likelihood classifiers are constructed from the following  event variables. The variables broadly characterize the energy deposition of the primary, the energy deposition of the secondaries, and the multiplicity and distribution of the secondaries. For all variables used in either classifier, the mutual correlation factors are in modulus less than 0.8.
\begin{itemize}

\item[]{\bf Energy deposition on the primary track} is the sum of the energy depositions in the TOF and tracker hits associated with the primary track, excluding the hit closest to the reconstructed vertex. {For a fully-stopped particle, this variable is approximately proportional to the kinetic energy.} 

\item[]{\bf Average energy deposition on primary track} is the energy deposition on the primary track, above, divided by the  number of hits. For events reconstructed with $0.5 <\beta <0.7$, fast $|Z| = 2$ particles exhibit a larger typical average energy deposition compared to correctly-reconstructed particles with $|Z| = 1$. Thus, high values of this variable are associated with fast $^4$He nucleus events. 

\item[]{{\bf Max over mean energy deposition} is the ratio of the highest primary-track energy deposition to the average of the remaining primary-track energy depositions, excluding the energy deposition closest to the reconstructed vertex. This variable probes the high-energy deposition expected as a particle slows to a stop, which is not observed for high-$\beta$ events.}

\item[]{\bf Primary truncated mean d{\em E}/d{\em x}} is the mean of the smaller half of the reconstructed \dedx values for the TOF and tracker hits associated with the primary track. \dedx for a given hit is the energy deposition normalized by the distance traveled in the detector. This variable identifies the typical \dedx for the particle at its initial $\beta$.  Removing hits with larger \dedx reduces the spread due to Landau fluctuations and due to the decrease in $\beta$ as the particle traverses the tracker. Fast primaries with $|Z| =1$ have lower values of this variable compared to antiprotons with $0.25<\beta<0.65$.

\item[]{\bf Primary TOF d{\em E}/d{\em x}} is the average \dedx of the outer and inner TOF hits associated with the primary track. {As illustrated in the upper right panel of Fig.~\ref{fig:vars},} this variable gives the cleanest representation of the $Z^2 /\beta^2$ energy deposition pattern prior to energy loss in the tracker. 

\item[]{{\bf TOF d{\em E}/d{\em x} over truncated mean d{\em E}/d{\em x}} is the ratio of primary TOF \dedx variable, above, to the truncated mean d{\em E}/d{\em x} variable, above. This variable rejects instrumental antiprotons based on the anomalously high energy deposition at the hard interaction point.  }

\item[]{{\bf Vertex energy over truncated mean d{\em E}/d{\em x}} is the ratio of the primary-track energy deposition closest to the reconstructed vertex to the truncated mean energy deposition. A large value is expected for slow particles that lose energy at a higher rate, and thus slow down faster, compared to fast particles.}

\item []{\bf Total energy deposition in the outer TOF} is the sum of all energy depositions from primary and secondary tracks in the umbrella and cortina. Large values of this variable correspond to instrumental antiprotons. 

\item[]{{\bf Energy deposition within 45\,cm of the vertex} is the total energy deposited in detectors within a sphere of radius 45\,cm from the reconstructed vertex. This variable  scales with the number and particle type of the secondary tracks, where a larger number of tracks results in a larger value.}

\item[]{\bf Average energy deposition per hit} is the sum of all energy depositions in the TOF and tracker divided by the total number of hits. {As illustrated in the lower right panel of Fig.~\ref{fig:vars},} this variable picks out the higher energy depositions of the $Z=2$ $^4$He background relative to $Z=1$ particles. Due to inclusion of the energy deposition associated with the vertex, it is not strongly correlated with the energy deposition on primary track variable.

\item[]{\bf Number of secondary tracks from the vertex} is the total number of reconstructed secondary tracks emerging from the reconstructed vertex. 
{As illustrated in the upper left panel of Fig.~\ref{fig:vars},} the typical multiplicity of secondary tracks is higher for antiproton annihilations compared to hard proton interactions.

\item[]{\bf Tracker number of hits} is the total number of energy depositions in the tracker. This variable scales with the number of secondary particles. 

\item[]{\bf Isotropy of secondary hits in the TOF cube} characterizes the degree to which the secondary tracks are boosted in the direction of the primary. 
{Each reconstructed secondary track $s$ emerges from the vertex with an angle $\phi_s$ relative to the primary, such that $\phi_s$ is defined by the primary momentum direction, the reconstructed vertex, and the hit in the TOF cube corresponding to track $s$. This variable is the mean  $\cos\phi_s$ over all secondary hits in the cube.} 
All hard interactions of positive nuclei occur in flight, resulting in a boosted event topology, while antinucleus annihilations may occur in flight or at rest. This variable is most useful for events with large numbers of secondary tracks.   

\item[]{\bf Isotropy of secondary hits in the tracker} is constructed similarly to the previous isotropy variable, but using all hits in the tracker rather than the TOF cube. This variable folds the isotropy of the secondary tracks with the depth of the vertex in the tracker, as a vertex deeper in the tracker results in relatively more hits for backward-going tracks. 

\item[]{\bf Average $\beta$ of secondary tracks} is calculated using the reconstructed time of the annihilation and the timestamps of the successive hits in the TOF cube.  {As illustrated in the lower left panel of Fig.~\ref{fig:vars},} antiproton annihilations typically result in faster secondaries compared to positive nucleus interactions. While antiprotons can annihilate entirely to pions, hard interactions of positive nuclei must produce heavier baryons to conserve baryon number. The distribution of this variable extends beyond ${\beta} = 1$ due to resolution effects. 

\end{itemize}

 In the likelihood analysis, probability distributions $P_i^a(q;\beta,\theta)$ of obtaining value $q$ for each event variable $i$ are first constructed for each event-type $a$ of interest using simulations. 
Then, in the analysis phase, all individual reconstructed events, are evaluated against all of the $P_i^a(q;\beta,\theta)$ to determine their probability of being of event-type $a$. 
$P_i^a(q;\beta,\theta)$ were  constructed bin-wise in true $\beta$ and $\theta$ for each event-type $a$  using simulated events and then smeared according to the $\beta$-dependent $\beta$ resolution.

\begin{table}[tbp]
\centering
\begin{tabular}{|l|cc|}
\hline
Variable&$\beta$ & ID\\
\hline
Energy deposition on the primary track&& \checkmark \\
Average energy deposition on primary track & \checkmark & \\
Max over mean energy deposition & \checkmark & \\
Primary Truncated mean d{\em E}/d{\em x} &\checkmark &\\
Primary TOF d{\em E}/d{\em x}&& \checkmark \\
TOF d{\em E}/d{\em x} over truncated mean d{\em E}/d{\em x} & \checkmark & \\
Vertex energy over truncated mean d{\em E}/d{\em x} & \checkmark & \\
\hline
Total energy deposition in the outer TOF & \checkmark & \\
Energy deposition within 45\,cm of the vertex& \checkmark & \\
Average energy deposition per hit&& \checkmark \\
\hline
Number of secondary tracks from the vertex & \checkmark &\checkmark \\
Tracker number of hits && \checkmark \\
Isotropy of secondary hits in the TOF cube&\checkmark& \checkmark \\
Isotropy of secondary hits in the tracker&& \checkmark \\
Average $\beta$ of secondary tracks&& \checkmark \\
\hline
\end{tabular}
\caption{\label{tab:vars} The event variables used in construction of the $\beta$-reconstruction ($\beta$) and identification (ID) likelihood classifiers characterize the energy deposition on the primary track (upper), the energy deposition of all of the particles (center), and the number and distribution of secondary tracks (lower).}
\end{table}

Table~\ref{tab:vars} indicates the event variables used in the construction of each likelihood classifier. 
Here, the event-types $a$ are  the antiproton ($\bar{p}$) signal or the proton ($p$) and $^4$He nucleus ($\alpha$) backgrounds. 
For the $\beta$-reconstruction likelihood, the $P_i^{\bar{p}}$ distributions were constructed using events reconstructed within $0.1$ of the true $\beta$ while the $P_i^{{p}}$ and $P_i^{\alpha}$ distributions were constructed using simulated events reconstructed >0.3 below the true $\beta$ or with true $\beta>0.8$. Probability distributions for use in the identification likelihood were calculated using all simulated events. 

{For a particular reconstructed event in the analysis, the probability that this event is of type $a$ is calculated using the values $q_i$ of its event variables as: 
\begin{equation}\label{eq:proba}
\mathcal{P}^a = \sqrt[N]{\prod_i^N P^a_i(q_i;\beta,\theta)}.
\end{equation}
The signal likelihood ratio $L$ is then calculated as
\begin{equation}\label{eq:llh}
L = \frac{\mathcal{P}^{\bar{p}}}{\mathcal{P}^{\bar{p}} + \mathcal{P}^{p}+ \mathcal{P}^{\alpha}}.
\end{equation}}
For both the $\beta$-reconstruction and identification likelihood, analysis is based on the natural logarithm of the ratio calculated in Eq.~\eqref{eq:llh}. A low value of $-\textrm{ln}(L)$ indicates high probability of a signal event. 

\begin{figure}[tb]
\centering
\includegraphics[width=0.48\textwidth]{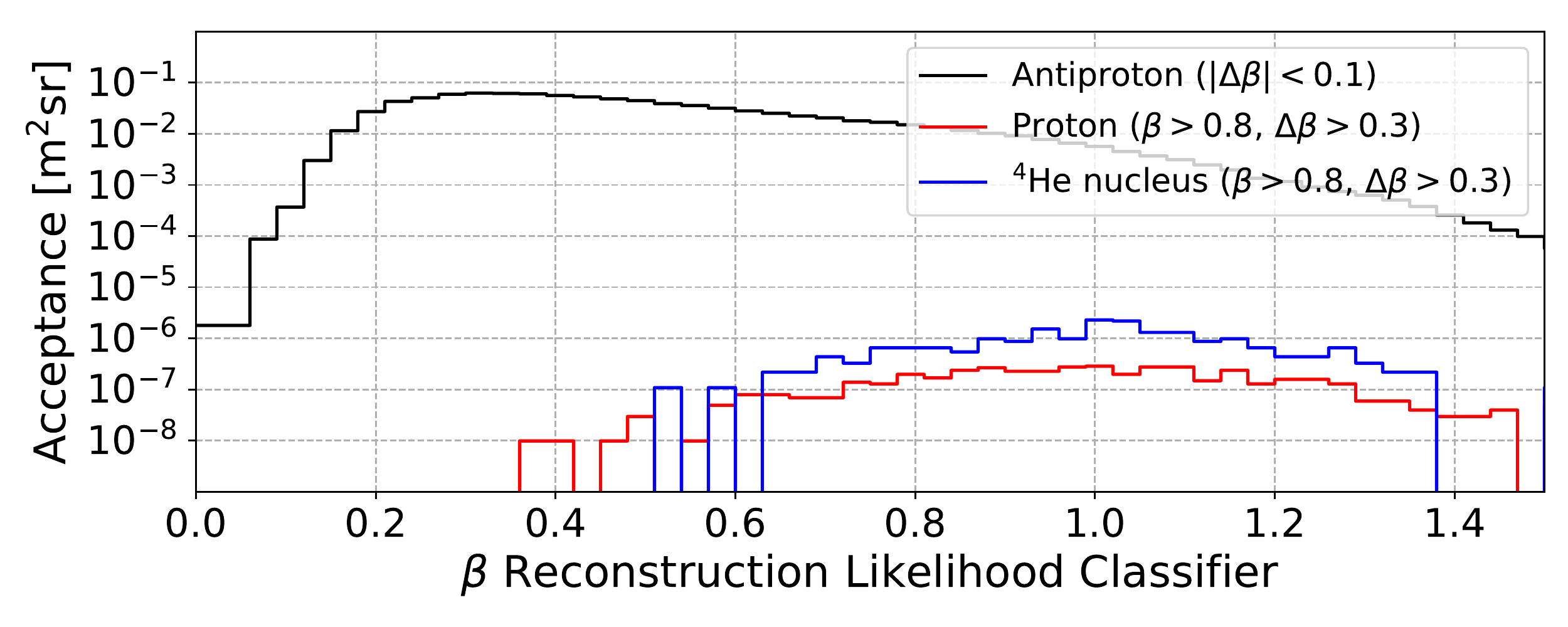}\caption{\label{fig:betallh} 
Distributions of the $\beta$-reconstruction likelihood classifier are shown for well-reconstructed antiproton events (black) as well as high-velocity proton events (red) and $^4$He nucleus events (blue) that were misreconstructed toward lower $\beta$. The distributions are shown for events reconstructed with $0.34\leq\beta<0.40$ and $\cos\theta > 0.75$ that have passed the trigger and preselection conditions. 
}
\end{figure}

\begin{figure}[tb]
\centering
\includegraphics[width=0.48\textwidth]{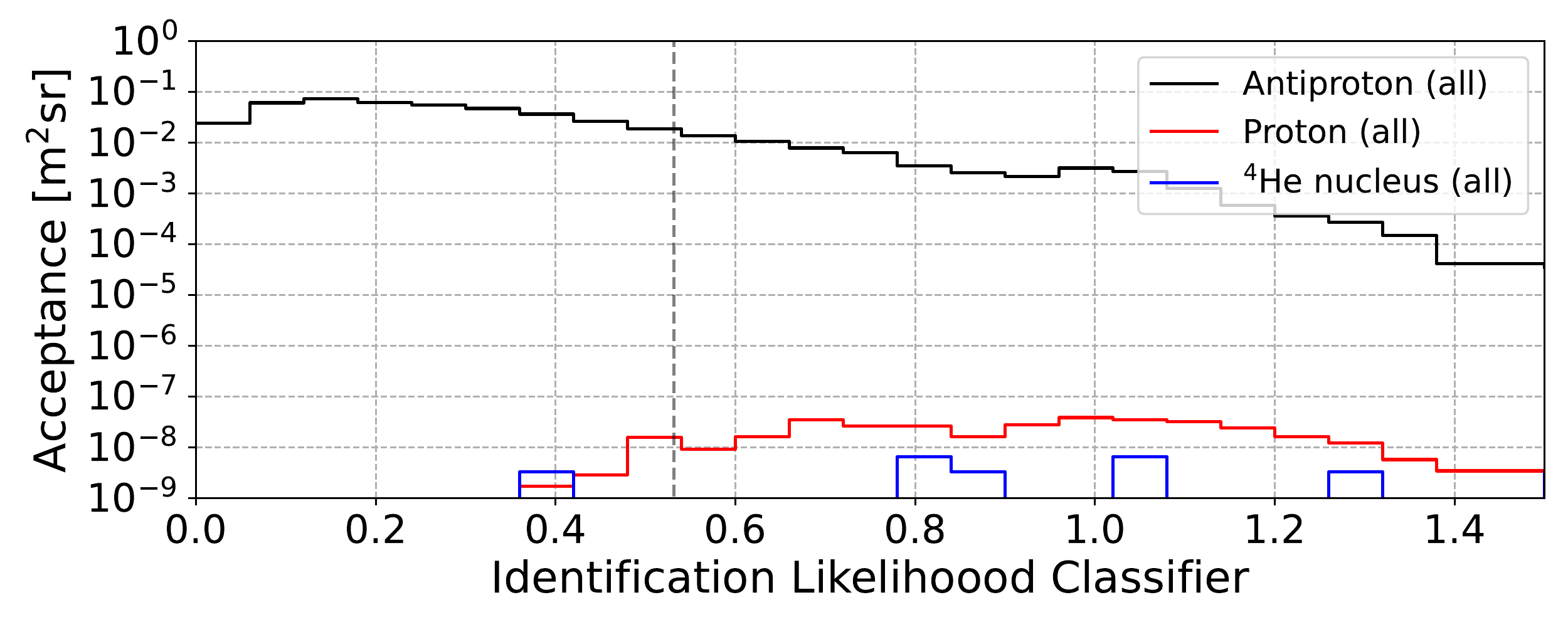}
\caption{\label{fig:llh} 
Distributions of the identification likelihood classifier are shown for  antiprotons (black), protons (red), and $^4$He nuclei (blue) reconstructed with $0.34\leq\beta<0.40$ and $\cos\theta > 0.75$. The distributions are shown for events that have passed trigger conditions, preselection, and the  $\beta$-reconstruction likelihood classifier cut.  Events in this analysis bin are selected if the identification likelihood classifier is less than 0.53 (gray dash). 
}
\end{figure}

Cuts are first applied based on the $\beta$-reconstruction likelihood classifier. 
Figure~\ref{fig:betallh} illustrates the distinct distributions of this classifier for well-reconstructed antiprotons as compared to high-$\beta$ positive nuclei
for the sample bin reconstructed with $0.34\leq\beta<0.40$ and $\cos\theta > 0.75$. The cut is constructed as a second-degree polynomial in the reconstructed $\beta$ to reject high-$\beta$ backgrounds while {maintaining a minimum required signal efficiency. The required efficiency is 50\% for events reconstructed with $\beta < 0.45$. At higher reconstructed $\beta$, the required efficiency decreases, down to 20\% at $\beta = 0.64$, where a stricter cut is required to reject the larger numbers of high-$\beta$ nuclei reconstructed in the higher part of the GAPS $\beta$ range. Depending on the reconstructed $\beta$, the resulting background rejection factor is in the range of $10^2 - 10^3$}.

Following the cut on the $\beta$-reconstruction likelihood classifier, cuts are applied based on the identification likelihood classifier. 
This analysis proceeds bin-wise in reconstructed $\beta$, and is conducted in two $\theta$ ranges: $0.5< \theta <0.75$ and $0.75 < \theta < 1.0$. For each bin, the target signal-to-background acceptance ratio is determined based on the TOI fluxes in Sec.~\ref{sec:fluxes} such that {subtraction of the proton and $^4$He nucleus contamination contributes a small statistical uncertainty compared to the irreducible atmospheric antiproton background}. 
Figure~\ref{fig:llh} illustrates the signal and background distributions and the optimized identification likelihood classifier cut for the bin with $0.34\leq\beta<0.40$ and $\cos\theta > 0.75$. In this bin, the  $^4$He nucleus acceptance has been reduced below the required level by the earlier steps in the analysis, including the charge cut. The required antiproton-to-proton acceptance ratio is meanwhile achieved by imposing a cut on the identification likelihood classifier.

\subsection{Calculating the Signal and Background Acceptance}\label{sec:accep}

The acceptance for each species is calculated based on the number of the simulated events passing all selection criteria, following \cite{Sullivan71}. Starting from the known geometric acceptance $\Gamma = 182\,\textrm{m}^{2}\textrm{sr}$ of the surface from which simulated events are generated, the final acceptance is proportional to the fraction of simulated events passing all analysis cuts. Here, the acceptance is analyzed by binning the simulated events according to the generated or reconstructed $\beta$ and $\cos\theta$.

\begin{figure}[tb]
\centering
\includegraphics[width=0.48\textwidth]{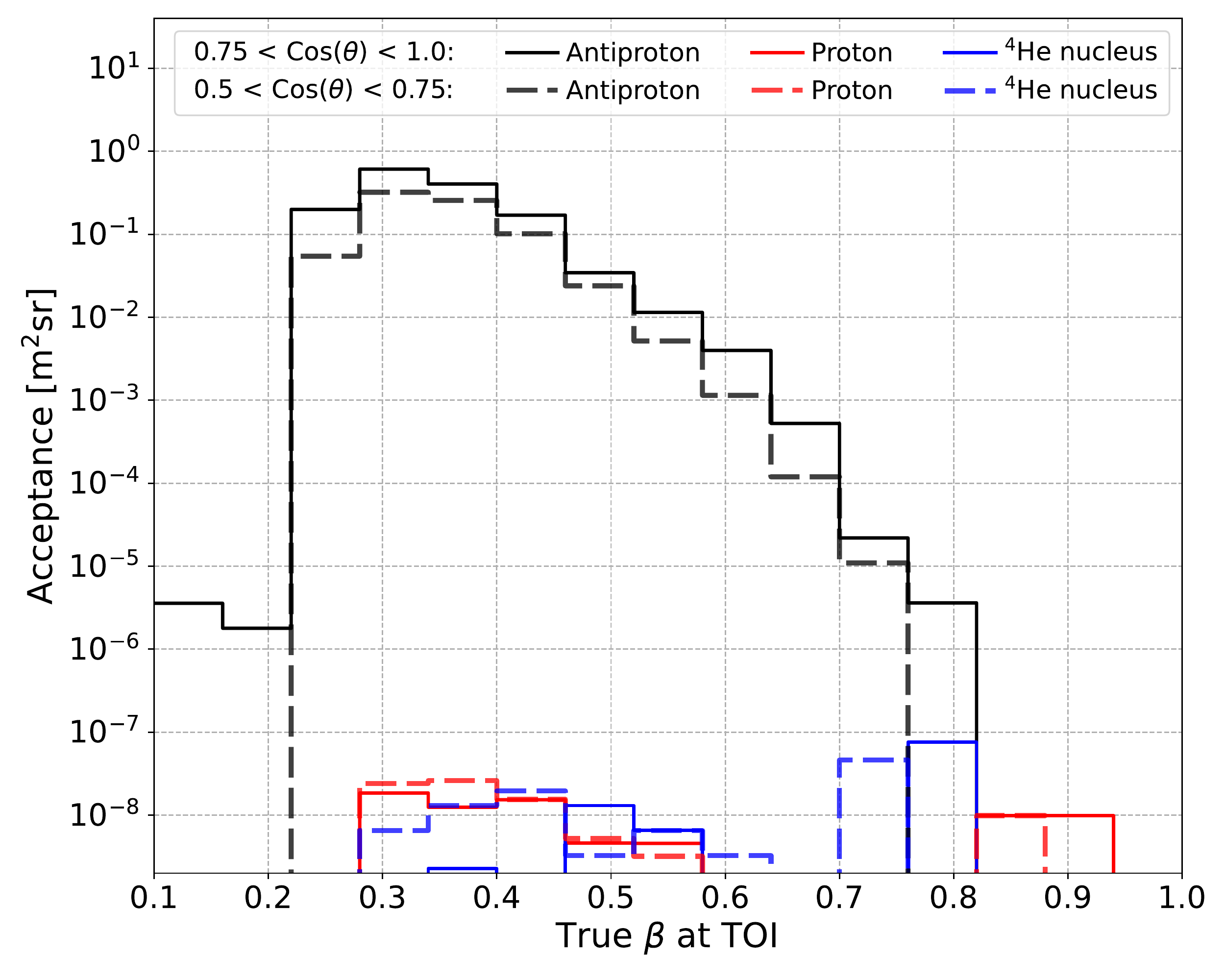}
\caption{\label{fig:accep}Acceptance of the GAPS instrument for antiprotons (black) as well as background species protons (red) and $^4$He nuclei (blue) is shown for events passing all selection criteria. Though the analysis is performed in terms of the reconstructed $\beta$, the acceptance is binned in the true $\beta$ at TOI. The acceptance is presented in two bins in $\cos\theta$. }
\end{figure}

Fig.~\ref{fig:accep} presents the resulting acceptance for each species. 
Both background species have been rejected at the target levels for a precision antiproton spectrum. 
While the analysis is performed using reconstructed information, the final acceptance is presented in terms of the true $\beta$ and $\cos\theta$ to facilitate comparison with the simulated fluxes.
The reconstructed $\beta$ is constrained to $0.25 < \beta <0.65$. The non-zero acceptance for events with true $\beta$ outside of this range is due to the intrinsic $\beta$ resolution.

\subsection{The Anticipated Antiproton Spectrum}\label{sec:toi}

\begin{figure}[tbp]
\centering
\includegraphics[width = 0.48\textwidth]{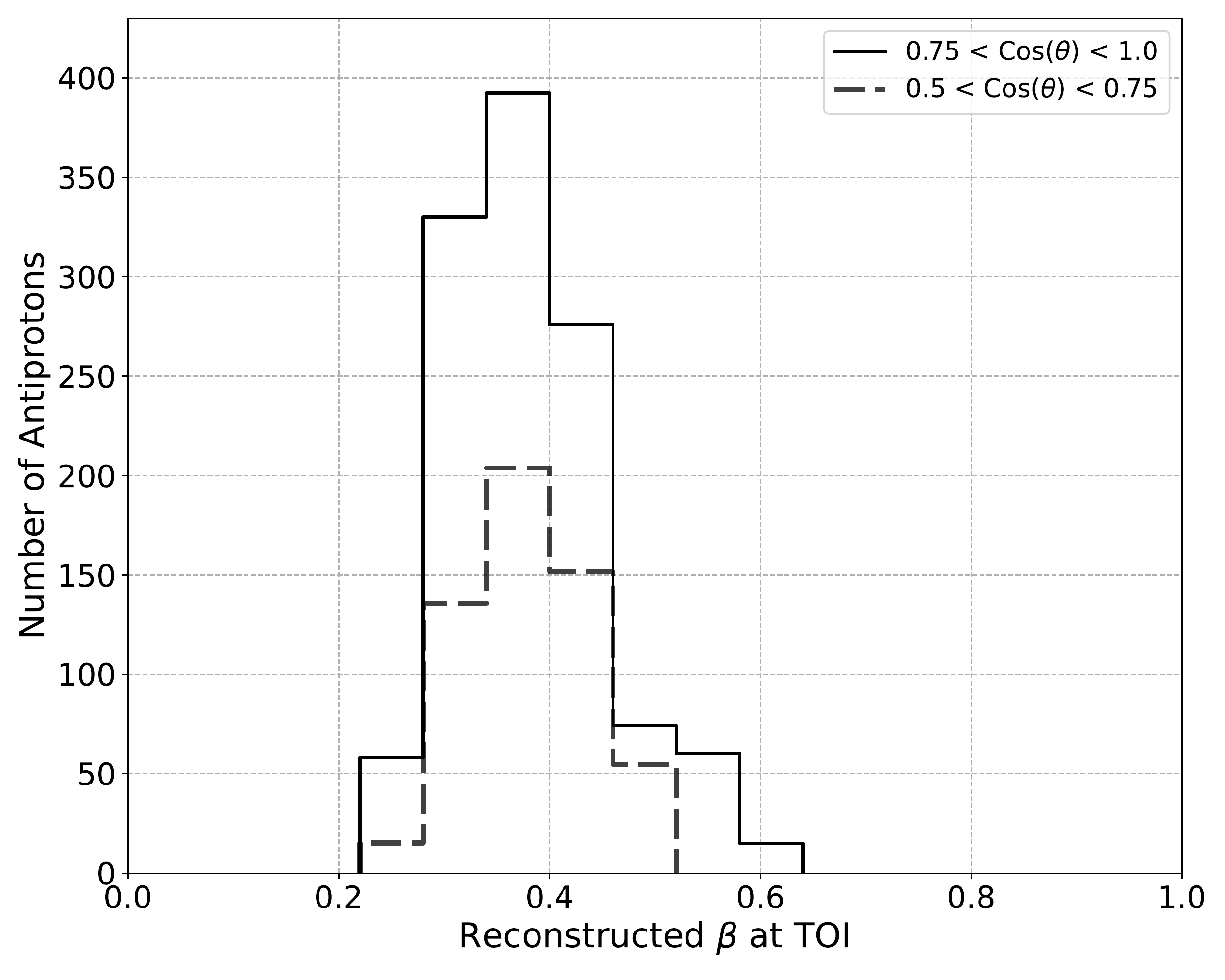}
\caption{\label{fig:toispec}{The total number of  antiprotons (cosmic + atmospheric) expected in three 35-day flights (90\% livetime) is shown for two ranges in $\cos\theta$. }
 }
\end{figure}

{Fig.~\ref{fig:toispec} illustrates the number of antiprotons (atmospheric + cosmic) expected in this analysis in three 35-day flights. 
The reported number is scaled based on the flux expected for the December 2022 solar activity. 
The bin width $\Delta\beta = 0.06$ is much larger than the resolution $\Delta\beta_{RMS}\leq$0.02 in the GAPS $\beta$ range.
The total number $N_a$ of  particles of each species $a$ is calculated using Eq.~\eqref{eq:accep}, considering the distribution in true $\beta$ for every reconstructed $\beta$ bin due to resolution effects. 
Assuming 90\% livetime, \totalTOIProgram~low-energy antiprotons are expected to be detected.} 
Meanwhile, \textasciitilde$40$ protons, \textasciitilde15 $^4$He nuclei, and even fewer other nuclei are expected to pass all selection criteria in the antiproton signal region.

{Translating the TOI measurement to a cosmic antiproton spectrum at TOA requires statistical subtraction of both the positive nuclei and atmospheric antiprotons expected in the signal region and correction for atmospheric losses. For each data point at TOA, atmospheric $\beta$ attenuation (Fig.\ \ref{fig:atmos}) is used to obtain the corresponding $\beta$ range at TOI for each $\theta$ bin. 
Then, the expected number of signal and background events reconstructed in the corresponding $\beta$ range at TOI is calculated. }

\begin{figure}[tbp]
\centering
\includegraphics[width = 0.48\textwidth]{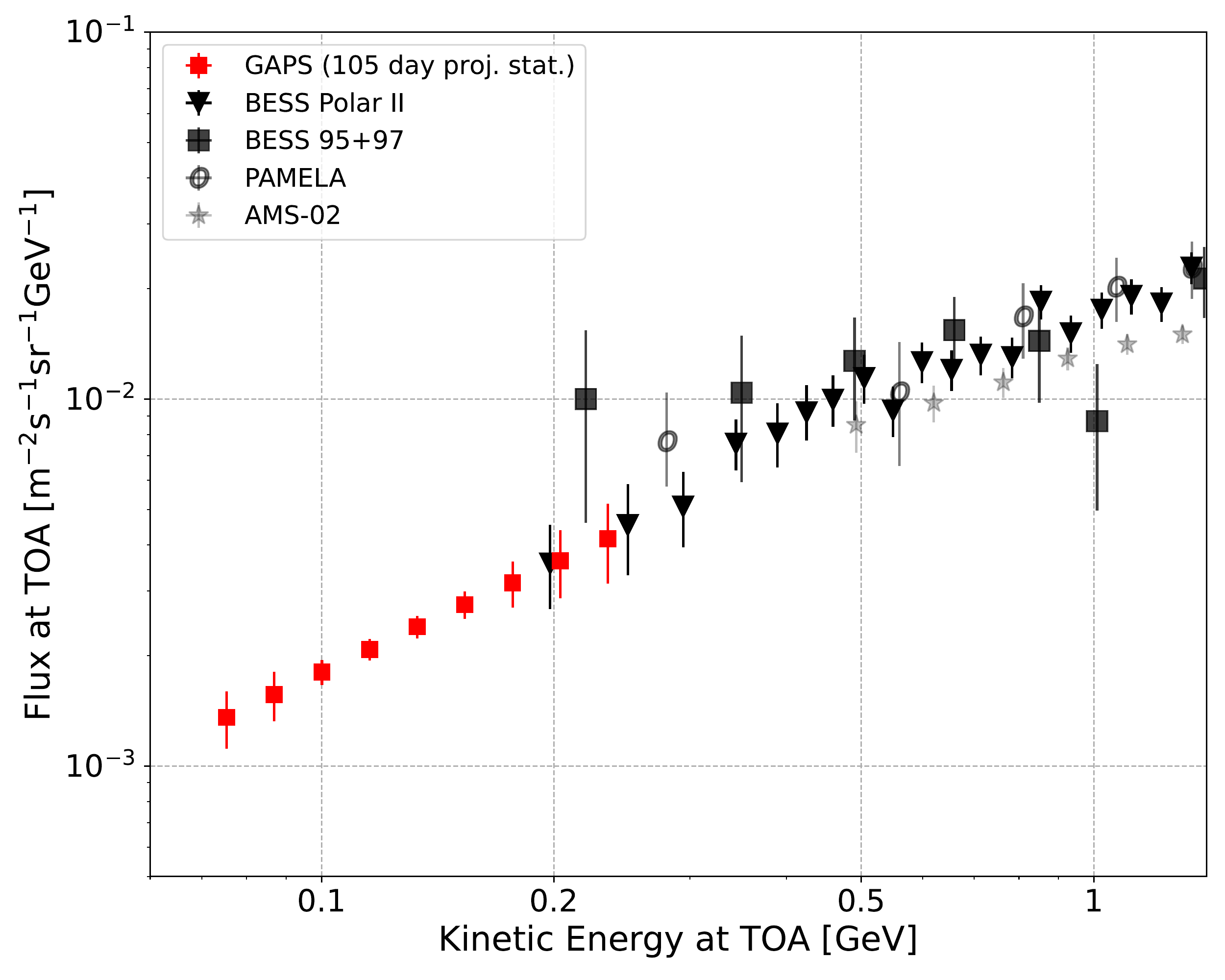}
\caption{\label{fig:spec}The projected GAPS precision cosmic antiproton spectrum (red) at the top of the atmosphere is shown with the statistics expected from three $35$-day flights. Data from BESS (1995 and `97 solar minimum; \cite{Orito20}), BESS Polar II (2007$-$08 solar minimum; \cite{BessPolarPbar}), PAMELA (2006$-$09 with \textasciitilde550\,MV best-fit solar modulation potential; \cite{PamelaPbar2}), and AMS-02 (2011$-$18 with average solar modulation potential \textasciitilde620\,MV;  \cite{AMSPbar,Aguilar21}) are also shown. }
\end{figure}

Fig.\ \ref{fig:spec} shows the resulting anticipated cosmic spectrum at TOA for three LDB flights. This is a naive scaling using the flux modeled for the Dec.~2022 solar conditions. Detection of \totalcosmicProgram~cosmic antiprotons is expected after subtraction of the atmospheric antiproton background. The error bars illustrate the expected 1$\sigma$ statistical uncertainty in the range of $6 - 25$\% per bin. 
Considering only data from the first LDB flight, the statistical errors will be larger by a factor of {$\sqrt{3}$}. 
This spectrum extends to lower energies than any previous cosmic antiproton measurement, with low statistical uncertainty. 
Expected sources of systematic error include the modeling of the atmospheric antiproton flux and of the atmospheric attenuation effects, the antiproton annihilation model, estimation of background nucleus contamination, and 
$\beta$ resolution effects. 
The atmospheric model will be calibrated to GAPS measurements of atmospheric antiprotons and deuterons to ensure a self-consistent result. 
$\beta$-resolution effects will be constrained based on ground and in-flight measurements of atmospherically-produced minimum-ionizing particles (e.g., atmospherically-produced muons) using a dedicated trigger. Work is ongoing to improve the modeling of antiproton-nucleus annihilation and its implementation in {\tt Geant4}.

\section{Conclusion and Outlook}\label{sec:conclusions}
This study uses a full instrument simulation with event reconstruction to demonstrate the power of the GAPS particle identification method for detecting cosmic antiprotons while rejecting cosmic-ray backgrounds. 
Antiprotons of both cosmic and atmospheric origin contribute to the flux at the anticipated 37\,km float altitude. 
In its first flight, GAPS will detect \totalTOI~antiprotons arriving with $0.25 < \beta < 0.65$ and $0.5 <\cos\theta <1.0$ at TOI. 
Using a standard model of Galactic propagation, solar and geomagnetic modulation, and atmospheric effects, this study shows that significant detection of \totalcosmic~cosmic antiprotons per flight is expected after subtraction of the atmospheric background. 
This corresponds to a high-statistics cosmic antiproton spectrum 
in the unprecedentedly low-energy range of \textasciitilde\range\,GeV/$n$ at TOA. 
Analysis of events arriving with $\cos\theta < 0.5$ is deferred to a future study. The flux with $0<\cos\theta < 0.5$ is dominated by atmospheric antiprotons, and while these events do not contribute to the cosmic antiproton spectrum, they will be critical to constrain systematic effects related to the atmospheric model.

{With this unprecedented statistical power in a never-before probed low-energy regime, the GAPS antiproton measurement will search for new physics including DM annihilation and local PBH evaporation. 
It will provide the first spectral data for comparison with Galactic and solar propagation models in a sensitive low-energy regime. 
This measurement will also validate the GAPS particle identification in flight, paving the way for rare-event searches with heavier antinuclei.}

Future developments in the analysis techniques to reject high-$\beta$ backgrounds are expected to further increase the sensitivity of the GAPS cosmic antiproton measurement. 
The background-rejection power of a slow-down fit assessing the compatibility of the reconstructed $\beta$ with the pattern of energy depositions on the primary track will be reported in a future publication.

\section*{Acknowledgment}

Funding: This work is supported in the U.S. by the NASA APRA program (Grant Nos.\ NNX17AB44G, NNX17AB46G, and NNX17AB47G), in Japan by the JAXA/ISAS Small Science Program FY2017, and in Italy by Istituto Nazionale di Fisica Nucleare (INFN) and the Italian Space Agency (ASI) through the ASI INFN agreement No.\ 2018-22-HH.0: ``Partecipazione italiana al GAPS - General AntiParticle Spectrometer.'' F.\ Rogers is supported through the National Science Foundation Graduate Research Fellowship Program under Grant No.\ 1122374. H.\ Fuke is supported by JSPS KAKENHI grants (JP17H01136, JP19H05198, and JP22H00147) and Mitsubishi Foundation Research Grant 2019-10038.  The contributions of C.\ Gerrity are supported by NASA under award No.\ 80NSSC19K1425 of the Future Investigators in NASA Earth and Space Science and Technology (FINESST) program. R.\ A.\ Ong receives support from the UCLA Division of Physical Sciences. K.\ Perez and M.\ Xiao are supported by Heising-Simons award 2018-0766. Y.\ Shimizu receives support from JSPS KAKENHI grant JP20K04002 and Sumitomo Foundation Grant No.\ 180322. M.~Yamatani receives support from JSPS KAKENHI grant JP22K14065. 

The technical support and advanced computing resources from University of Hawaii Information Technology Services – Cyberinfrastructure, funded in part by the National Science Foundation MRI Grant No.\ 1920304, are gratefully acknowledged. This research was performed using resources provided by the Open Science Grid \cite{osg07,osg09}, which is supported by the National Science Foundation Grant No.\ 2030508.

\bibliographystyle{naturemag_noURL}
\bibliography{main}

\end{document}